\documentclass[proceedings]{JHEP} 

\usepackage{epsfig}                     

\newbox\mybox
\newcommand\fverb{\setbox\mybox=\hbox\bgroup\verb}
\newcommand\fverbdo{\egroup\medskip\noindent\fbox{\unhbox\mybox}\ }
\newcommand\fverbit{\egroup\item[\fbox{\unhbox\mybox}]}

\font\beeg=cmr17 scaled 1600            
\newcommand\init[1]{\setbox\mybox=\hbox{{\beeg #1}~}%
                   \noindent\global\hangindent=\wd\mybox\global\hangafter-2%
                   \sc\smash{\llap {\lower 13.2pt \box\mybox}}}

\newdimen\unit
\def\point#1 #2 #3{\vbox to0pt{\kern-#2\unit
  \hbox{\kern#1\unit#3}\vss}
 \nointerlineskip}

\def\la{\langle}
\def\ra{\rangle}
\def\eps{\varepsilon}

\def\l{\left}
\def\r{\right}
\def\nn{\nonumber}

\def\ord#1{{\mathcal O}\l(#1\r)}

\def\beq{\begin{equation}}
\def\eeq{\end{equation}}
\def\bea{\begin{eqnarray}}
\def\eea{\end{eqnarray}}

\def\eq#1{Eq.\ (\ref{#1})}

\def\tab#1{Table \ref{#1}}

\def\sec#1{Section \ref{#1}}

\def\fig#1{Figure \ref{#1}}

\def\mev{\mbox{ MeV}}
\def\gev{\mbox{ GeV}}

\def\gsim{\vbox{\vspace{4pt}\hbox{ $\stackrel{>}{\sim}$ }\vspace{-4pt}}}
\def\lsim{\vbox{\vspace{4pt}\hbox{ $\stackrel{<}{\sim}$ }\vspace{-4pt}}}

\title{Exclusive semileptonic $B$ decays: lattice results and
  dispersive bounds\footnote{CERN-TH/99-352, LAPTH-Conf-765/99,
    CPT-99/P.3911.}}

\author{Laurent Lellouch\thanks{Address from October, 1999: LAPTH,
        Chemin de Bellevue, B.P. 110, F-74941 Annecy-le-Vieux Cedex,
         France.  On leave from: Centre de
     Physique Th\'eorique, Case 907, CNRS Luminy, F-13288 Marseille
     Cedex 9, France.}\\
        Theory Division, CERN, CH-1211 Geneva 23, Switzerland\\
        E-mail: \email{laurent.lellouch@cern.ch}}

\conference{Heavy Flavours 8, Southampton, UK, 1999}

\abstract{The current status of lattice and dispersive bound
  calculations for exclusive semileptonic $B$ decays is reviewed.
  Emphasis is placed on decays relevant for the measurement of the 
  sides of the unitarity triangle determined by $|V_{ub}|$ and
  $|V_{cb}|$. }

\begin{document}

\maketitle 


\section{Introduction}

Exclusive semileptonic decays of hadrons containing a $b$ quark play
an important role in testing the Standard Model. They enable the
measurement of two of the three sides of the unitarity triangle,
determined by the Cabibbo-Kobayashi-Maskawa (CKM) parameters
$|V_{ub}|$ and $|V_{cb}|$.  However, such measurements require a
quantification of the non-perturbative, strong interaction dynamics
which modify the elementary coupling of the $b$ and resulting quark to
the $W$ boson.  Lattice QCD and dispersive bounds provide tools to
quantify and constrain these non-perturbative effects from first
principles\footnote{Lattice results relevant for leptonic decays,
  neutral $B$-meson mixing and related topics, also useful for
  constraining the unitarity triangle, are covered in Shoji
  Hashimoto's talk~\cite{SHa99}.}. Model independence is crucial for
the unitarity-triangle tests of the Standard Model to be meaningful.
Without it, it will be impossible to disentangle potential new physics
effects from artefacts of the models used.

The talk is divided into two main parts: lattice results and
dispersive bounds. I will begin the first part with a brief
introduction to the systematic uncertainties of lattice calculations
and to the different ways in which heavy quarks can be studied on a
discrete spacetime. I will then turn to lattice results relevant for
heavy-to-light quark decays, including $B\to\pi\ell\nu$,
$B\to\rho\ell\nu$ and $B\to K^*\gamma$. This will be followed by a
discussion of heavy-to-heavy quark decays, focussing mainly on $B\to
D^{(*)}\ell\nu$ decays\footnote{Other recent lattice reviews of some
  of the subjects covered here can be found in
  \cite{Hashimoto:1999bk,Flynn:1997ca}. The last Heavy Flavour review
  of lattice results for semileptonic decays is \cite{Flynn:1997ui}.}. The
second part will begin with a brief description of the methodology of
dispersive bounds and its applications to heavy-to-heavy semileptonic
decays. It will continue with applications of dispersive bound
techniques to heavy-to-light decays.

\section{Lattice Results}

Although lattice QCD provides a means of determining
non-perturbative, strong-interaction effects to arbitrary accuracy
from first principles, in practice the results suffer from a variety
of uncertainties due to limitations in computing resources. The main
ones are:

$\bullet${\it Statistical errors,} associated with the fact
that the QCD path integral is evaluated using Monte-Carlo simulations.
They are estimated with standard statistical techniques.

$\bullet${\it Discretisation errors,} associated with the
fact that spacetime is a discrete mesh of points. They are
particularly important for heavy quarks and are dealt with in a
varieties of ways, as discussed in \sec{sec:hlhads}.

$\bullet${\it Finite volume errors,} associated with the fact that in
lattice calculations, spacetime is a finite box. They are not
significant for the quantities discussed here.

$\bullet${\it Uncertainties associated with extrapolations in light
  quark masses}. On present day lattices, light quarks ($u$ and $d$)
must have masses larger than about $m_s/2$ so that the associated
pions do not feel the edges of the box. One must therefore
extrapolate results from these larger masses to the physical $u$ and $d$ 
quark masses.

$\bullet${\it Matching errors,} associated with the matching of
lattice results onto the continuum. This matching is necessary because
ultraviolet modes are treated differently on the lattice than they are
in continuum regularisations. While it can be performed
perturbatively, with resulting perturbative uncertainties, more and
more it is carried out non-perturbatively, eliminating it as a source
of systematic error. This is the case in most
non-perturbatively-improved calculations discussed below.

$\bullet${\it Quenching errors,} associated with the fact that in most
calculations of semileptonic $B$ decays, the feedback of quark loops
on the gauge fields is neglected. It is not anticipated that these
effects be more than $\ord{10\%}$ here. They have only very recently
begun to be taken into account for these decays \cite{Bernard:1999ic}.

\subsection{Heavy-light hadrons on the lattice}
\label{sec:hlhads}

Because the $b$ quark with mass $m_b\sim 5\gev$ has a compton
wavelength which is small compared to typical lattice spacings, $a\sim
(2-3\gev)^{-1}$, it cannot be simulated directly as a relativistic
quark on present day lattices. This has led to a variety of
approaches, which we review now for the case of hadrons composed of a
heavy quark and light degrees of freedom.

\subsubsection{Relativistic quarks}

Relativistic fermions are obtained from a discretisation of the
euclidean Dirac action. For heavy quarks, the discretisation most
commonly used is that of Wilson.  In modern calculations, it is often
$O(a)$-improved by use of a Shekholeslami-Wohlert (SW) action. This
means that for heavy quarks $Q$ of mass $m_Q$, discretisation errors
are formally reduced from $\ord{am_Q}$ to $\ord{\alpha_sam_Q}$ if the
improvement is performed at tree level, and to
$\ord{(\alpha_sam_Q)^2}$ if it is non-perturbative.

Despite these improvements, simulating the $b$ quark directly would
lead to uncontrollable discretisation errors. Therefore, what is done
is to perform the calculations for a number of quark masses around
that of the charm, where discretisation errors are under control, and
then extrapolate the results to $m_b$ using Heavy Quark Effective
Theory (HQET) as a guide. Typically, HQET predicts that a 
form factor $F$ will scale with $m_Q$ as
\beq
F(w)m_Q^\nu=A(w)\l(1+\frac{B(w)}{m_Q}+\ord{\frac{1}{m_Q^2}}\r)
\ ,
\label{eq:hqscal}
\eeq
at fixed four-velocity recoil $w\sim 1$, up to calculable logarithmic
corrections. Here, $\nu$, $A$ and $B$ depend on the form factor.

The main problem with this approach is this rather long extrapolation to
$m_b$.

\subsubsection{Effective theories}

The point of view here is that for heavy quarks whose mass $m_Q$ is large
compared to typical QCD scales, $\mu_{QCD}$, an expansion of QCD in powers
of $\mu_{QCD}/m_Q$ may be useful. Indeed, it reduces discretisation errors
from powers of $am_Q$ to powers of $a\mu_{QCD}$, the latter being small
on present day lattices.

{\it Static quarks.} The first implementation of this idea was to
consider heavy quarks as static, spin-$1/2$ color sources. The problem
with this approach is that accurate results for the physical $b$ quark
require that $1/m_b$-corrections be taken into account. This leads to
power divergences proportional to inverse power of the lattice spacing
which are difficult to subtract. It also leads to a proliferation of
operators whose matrix elements must be computed. It has the further
disadvanted of yielding correlators with poor signal to noise ratios.

{\it ``Non-relativistic QCD (NRQCD)''}, in the context of heavy-light
mesons, essentially corresponds to keeping in the action some of the
leading $1/m_Q$-corrections mentioned above, thereby implicitly
re-summing to infinite order their effects on the processes studied.
The problem with this approach is again power divergences and operator
proliferation.

In the {\it hybrid or ``Fermilab'' approach}, the goal is to find an
action valid for all $am_q$. In practice, it is a relativistic SW
action that is used and the results are subsequently given a
non-relativistic interpretation. There is debate over the extent to
which systematics are controlled in this approach.

\subsection{Heavy-to-light decays}

In heavy-to-light-quark semileptonic decays, the light, final state
hadron can have momenta as large as $|\vec{p}|\sim m_Q/2$ in the
parent rest frame. For $m_Q=m_b$, and on present day lattices, such
momenta would lead to uncontrollably large discretisation effects
proportional to powers of $a|\vec{p}|$. Therefore, at present, only a
limited kinematical range near the zero-recoil point can be reached
without extrapolation. Even so, lattice calculations are useful, for
the relevant matrix elements are not normalised by heavy quark
symmetry as they are for heavy-to-heavy quark decays (at zero recoil).
Furthermore, experiment is beginning to measure the corresponding
differential rates within the lattice's kinematical reach
\cite{Behrens:1999vv}, which will allow model-independent
determinations of CKM parameters such as $|V_{ub}|$.  Finally, lattice
groups are investigating new ways of exploring the region of large
recoils.

\subsubsection{$B^0\to\pi^-\ell^+\nu$}
\label{sec:bpi}

In the past two years, most studies of heavy-to-light decays have
concentrated on $B^0\to\pi^-\ell^+\nu$. The relevant matrix element is
$$
\la\pi^-(p')|V^\mu|B^0(p)\ra = \frac{M_B^2-M_\pi^2}{q^2}q^\mu f_0(q^2)
$$
\beq
+(p+p'-\frac{M_B^2-M_\pi^2}{q^2}q)^\mu f_+(q^2)
\ ,
\label{eq:bpiff}
\eeq
where $q=p-p'$ and $V^\mu=\bar b\gamma^\mu u$.

\FIGURE{\offinterlineskip
\hbox{\hfill\vbox{\epsfxsize=6.5cm\epsffile[100 210 530 620]{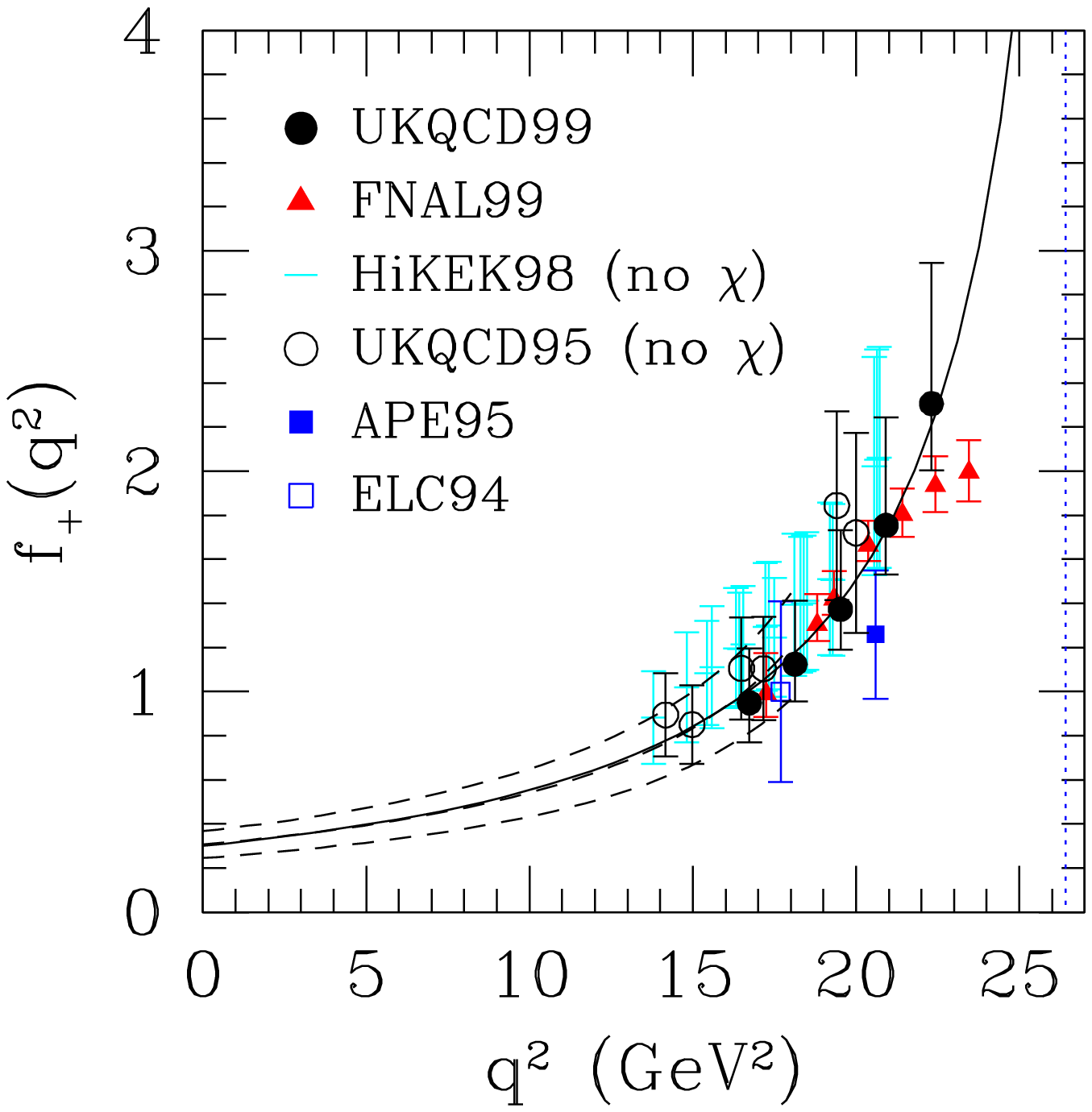}}
\vbox{\epsfxsize=6.5cm
\epsffile[100 210 530 620]{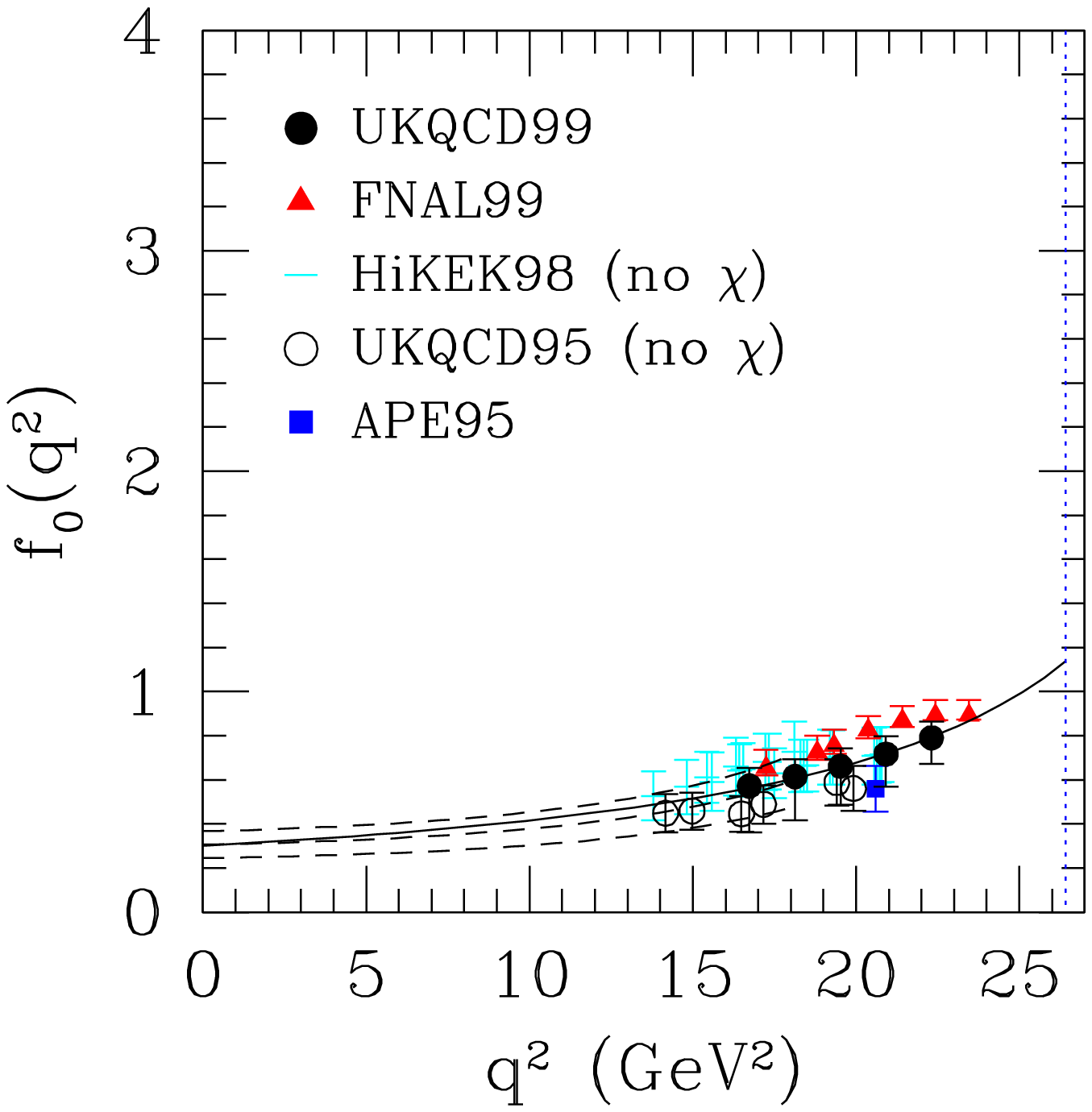}}\hfill}
\caption{Form factors for $B^0\to\pi^-\ell^+\nu$. The points are results
  from various lattice groups. The dashed lines are LCSR results (see
  text).  The solid lines are the result of the fit of UKQCD99's
  central value results, with statistical errors only, to the BK
  parameterisation of \eq{eq:bpiparam} \cite{Bowler:1999xn}.}
\label{fig:fpf0}
}
In \fig{fig:fpf0}, I have collected results for both $f_+$ and $f_0$.
Results from ELC94 \cite{Abada:1994dh}, APE95 \cite{Allton:1995ui},
UKQCD95 \cite{Burford:1995fc} and UKQCD99 \cite{Bowler:1999xn} were
obtained using relativistic quarks. APE95 and UKQCD95 implement
tree-level improvement, while UKQ\-CD99 implements full
non-perturbative improvement. The FNAL99
\cite{Ryan:1998tj,Ryan:1999kx} results were obtained with the hybrid
approach while HiKEK98 \cite{Hashimoto:1998sr} use NRQCD.  Besides the
non-perturbative improvement of UKQCD99 and the fact that FN\-AL99 has
begun exploring cutoff dependence by performing their calculation at
three values of the lattice spacing\footnote{The FNAL99 results shown
  here are those obtained at $\beta=5.9$.}, another novelty is the
fact that FNAL99 and UKQCD99 were able to extrapolate the form factors
in light quark mass to the physical $u$ and $d$ masses. APE95 and
ELC94 had only performed this extrapolation for a single $q^2$. As
indicated by the ``no $\chi$'', HiKEK98 and UKQCD95 have $u$ and $d$
quarks with masses around that of the strange.  A systematic
uncertainty has been added to these results to account for this fact,
as described in \cite{Lellouch:1996yv}. It should be noted that
UKQCD99's error bars include a wide range of systematics. Their
results are collected in \tab{tab:fpf0rate}.
\TABLE[t]{
\begin{tabular}{cccccc}
\hline
$q^2$ $($GeV$)^2$ & $16.7$ & $18.1$ & $19.5$ & $20.9$ & $22.3$ \\
\hline
$f_+(q^2)$ & $0.9^{+1}_{-2}\ {}^{+2}_{-1}$ 
        & $1.1^{+2}_{-2}\ {}^{+2}_{-1}$ 
        & $1.4^{+2}_{-2}\ {}^{+3}_{-1}$ 
        &$1.8^{+2}_{-2}\ {}^{+4}_{-1}$
        & $2.3^{+3}_{-3}\ {}^{+6}_{-1}$ \\     
$f_0(q^2)$ & $0.57^{+6}_{-6}\ {}^{+\ 5}_{-20}$
        & $0.61^{+6}_{-6}\ {}^{+\ 6}_{-19}$ 
        & $0.66^{+5}_{-5}\ {}^{+\ 6}_{-17}$ 
        & $0.72^{+5}_{-4}\ {}^{+\ 6}_{-14}$ 
        & $0.79^{+5}_{-4}\ {}^{+\ 6}_{-11}$ \\
$1/|V_{\rm ub}|^2{\rm d}\Gamma / {\rm d}q^2$ $({\rm ps}^{-1}{\rm GeV}^{-2})$
        & $0.29^{+10}_{-\ 9}\ {}^{+11}_{-\ 6}$
        & $0.27^{+8}_{-7}\ {}^{+11}_{-\ 1}$ 
        & $0.25^{+7}_{-6}\ {}^{+11}_{-\ 1}$
        & $0.23^{+6}_{-5}\ {}^{+11}_{-\ 1}$ 
        & $0.20^{+5}_{-5}\ {}^{+\ 9}_{-\ 1}$\\
\hline
\end{tabular}
\caption{Form factors and differential decay rate as functions of $q^2$ from
UKQCD99 \cite{Bowler:1999xn}.}
\label{tab:fpf0rate}
}

Despite the wide range of approaches, results for both form factors are
generally consistent. This is rather reassuring given the number of
fits and extrapolations involved. Also plotted in \fig{fig:fpf0} are
the light-cone sumrule (LCSR) results of \cite{Ball:1998tj} (see also
\cite{Belyaev:1993wp,Belyaev:1995zk,
  Khodjamirian:1997ub,Bagan:1997bp,Khodjamirian:1998vk}), including
a 20\% error band. In the region of overlap, agreement is
excellent.  Furthermore, the LCSR curves do look like natural
extensions of the lattice results.

The differential decay rate is easily obtained from the form factors,
up to an overall factor of $|V_{ub}|^2$. Therefore, comparison with
experiment in the range of pion momenta reached by the lattice
calculation will yield a model-independent determination of
$|V_{ub}|$. FNAL suggest to compare $d\Gamma/d|\vec{p}_\pi|$ in the
range $0.4\gev\le |\vec{p_\pi}|\le 0.8\gev$, where their systematic
errors are minimised \cite{Ryan:1998tj,Ryan:1999kx}. Their rate is
shown in \fig{fig:fnalrate}, together with that of UKQCD99. Agreement
is good, a bit less so at the lowest value of $|\vec{p_\pi}|$,
where sensitivity to the $B^*$ pole is strongest.  The results of
UKQCD99 for $d\Gamma/dq^2$ for $0.5\gev$ $\le |\vec{p_\pi}|\le
1.1\gev$, with an estimate of systematic errors,
are given in \tab{tab:fpf0rate}. 
\FIGURE{ \epsfxsize=6cm\epsffile[100 210 530 620]{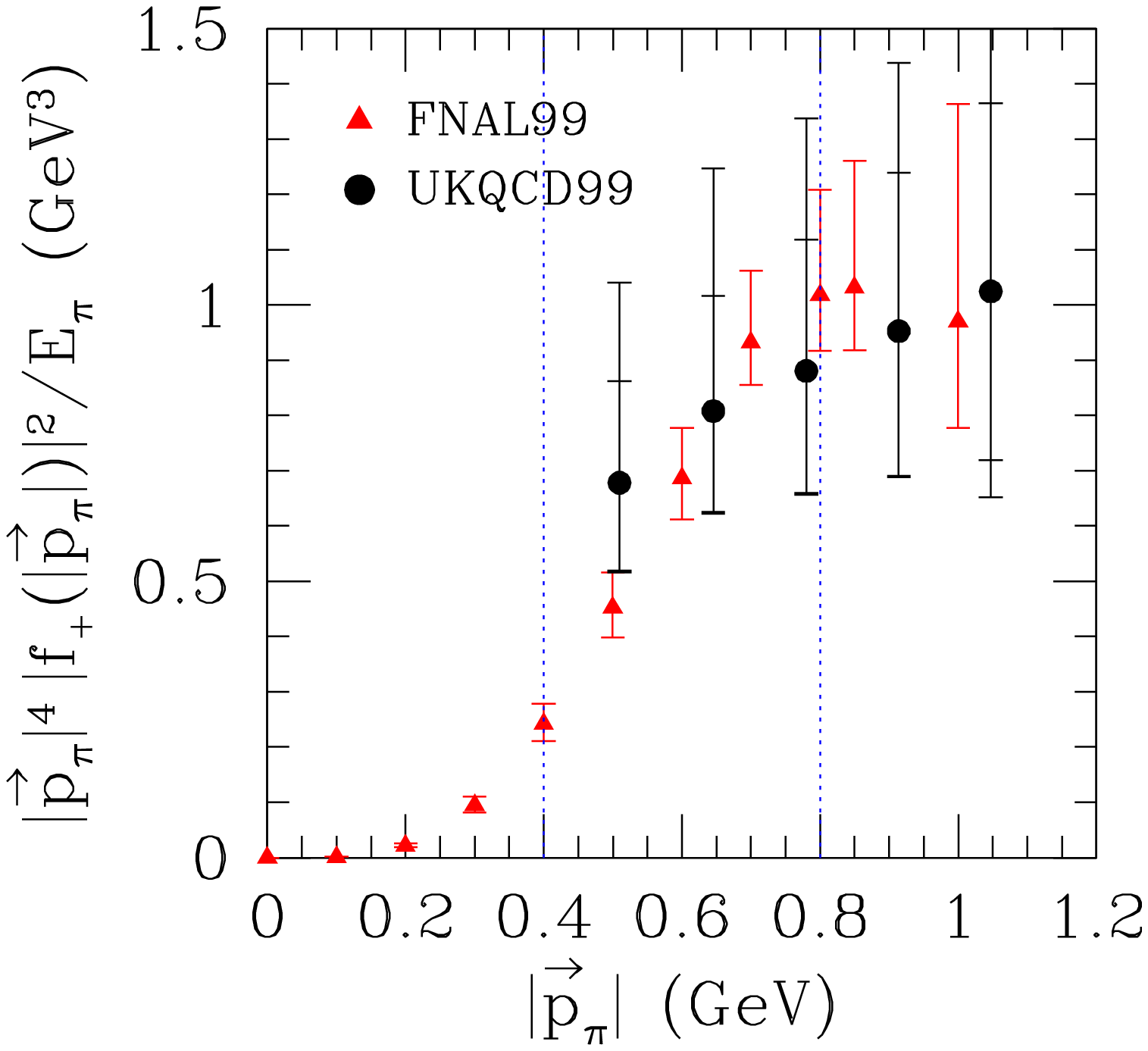}
\caption{Differential decay rate for 
  $B^0\to\pi^-\ell^+\nu$ from FNAL99 \cite{Ryan:1999kx} and
  UKQCD99 \cite{Bowler:1999xn}. The smaller error bars on the UKQCD99
  results are statistical; the larger ones are statistical and
  systematic combined in quadrature.}
\label{fig:fnalrate}}

\medskip

One may also attempt to extrapolate the a\-bove lattice results to the
large recoil region. While it can be done in a model-independent way
as described in \sec{sec:bpidisp}, extrapolation of the most recent
results has only been performed using models.  It is not, therefore,
on the same firm theoretical ground as the unextrapolated results.
Nevertheless, it may be useful if one needs information about the
decay over the full kinematical range\footnote{Please see
  \cite{Burford:1995fc,DelDebbio:1997kr} for earlier constrained
  $q^2$-extrapolations.}.
UKQCD99 has considered fits to
\bea f_+(q^2)&=&\frac{f(0)} {(1-q^2/M^2_{B^{\star}})(1-\alpha
  q^2/M^2_{B^{\star}})}\nn\\ f_0(q^2)&=&\frac{f(0)}{(1- q^2/\beta
  M^2_{B^{\star}})} \ ,\label{eq:bpiparam}
\eea
with either $\alpha=1$ and $f(0)$, $\beta$ as fit parameters
(dipole/pole), or leaving all three parameters free, as suggested by
Becirevic and Kaidalov (BK) \cite{Becirevic:1999kt}.  These two
parameterisations are consistent with the kinematical constraint
$f_+(0)=f_0(0)$, heavy-quark scaling at large $q^2$ (\eq{eq:hqscal}) 
and light-cone (LC) scaling at small $q^2$ (\eq{eq:lcscal})
\cite{DelDebbio:1997kr,Becirevic:1999kt}.  The BK parameterisation is
more physical in that it correctly takes into account the $B^*$ pole
contribution to $f_+$. It is important to use parameterisations which
are consistent with as many model-independent constraints as possible.

The result of UKQCD99's fit to the BK parameterisation is shown in
\fig{fig:fpf0}. Agreement with the LCSR results is stunning. UKQCD99
then uses its fits to obtain the total rate.  Their result, along with
older lattice results, is given in \tab{tab:f0rate}.  
Agreement is good, though error bars are large.
\TABLE{
\begin{tabular}{rcll} 
\hline
 & Rate  & $f_{+,0}(0)$ & description\\
\hline
UKQCD99 \cite{Bowler:1999xn}& $9^{+3+3}_{-2-4}$ & $0.30^{+6+4}_{-4-9}$ 
& {\small BK fit plus many systematics}\\
APE99 \cite{Abada:1999xd} & & 0.28(4) 
&{\small LC scaling at $q^2=0$}\\
UKQCD98 \cite{DelDebbio:1997kr} & $8.5^{+3.3}_{-1.4}$ & 0.27(11) 
& {\small dipole/pole fit to $f_+$/$f_0$}\\
APE95 \cite{Allton:1995ui}& $8\pm 4$ & 0.35(8)
& {\small $q^2{\simeq} 20.4\gev^2$, pole fit, $m_\mathrm{p}{=}5.32(1)\gev$}\\
ELC94 \cite{Abada:1994dh} & $9\pm 6$ & 0.30(14)(5) 
& {\small $q^2{\simeq} 18\gev^2$, pole fit, $m_\mathrm{p}{=}5.29(1)\gev$}\\
\hline
\end{tabular}
\caption{$f_{+,0}$ at $q^2=0$ and total rate in units of 
  $|V_{ub}|^2{\mathrm ps}^{-1}$.  The results of APE and ELC are
  obtained by extrapolation of $f_+$ determined at a single $q^2$
  value, as indicated in column 4.  UKQCD uses results for $f_+$ and
  $f_0$ at a number of $q^2$ values and performs constrained fits to
  the parameterisations of \eq{eq:bpiparam}. The final state
  pion in the UKQCD98 results is composed of quarks with masses around
  that of the strange and include a systematic error to account for
  this. The APE99 results was obtained as described around
  \eq{eq:lcscal}.}
\label{tab:f0rate}}

While the extrapolations discussed above introduce model-dependence,
there are many possible checks and constraints. For instance, the
residue of the $B^*$ pole in $f_+$ is related to the coupling of the
pion to heavy mesons, defined in \eq{eq:gbbstpidef}:
\beq 
f_{\mathrm{res}}\equiv \frac{\mathrm{Res}_{q^2=m_{B^*}^2}
\,f_+(q^2)}{M_{B^*}^2} =
\frac{g_{B^*B\pi}}{2f_{B^*}} \ ,\label{eq:residue}
\eeq
where $f_{B^*}$ is given by $\la 0|V_\mu|B^*\ra$ $=\eps_\mu
M_{B^*}^2/f_{B^*}$. Thus, to the extent that the BK parameterisation
is a valid description of $f_+$ and $f_0$, the direct determination of
this residue, $f_{\mathrm{res}}=$ $f(0)/(1-\alpha)$, must agree with
its indirect determination through $g_{B^*B\pi}$ and $f_{B^*}$. The
results of UKQCD99 yield $f_{\mathrm{res}}=0.55^{+7+51}_{-7-0}$ while
\eq{eq:residue} with $f_{B^*}=27^{+3+0}_{-3-5}$ from
\cite{Becirevic:1998ua} and $g$ from \eq{eq:gres} give
$f_{\mathrm{res}}=0.63^{+10+17}_{-10-12}$. Error bars are large and
$g$ from \eq{eq:gres} is exploratory and was converted to
$g_{B^*B\pi}$ using \eq{eq:gdef} without $1/m_b$-corrections.
Agreement within statistical errors, though, indicates a certain
consistency of the BK fit to the central values.  Furthermore,
statistical and systematic uncertainties on the indirect determination
are comparable to or smaller than those on the direct determination.
This suggests that some gain may result from constraining the $B^*$-pole
residue with its indirect value. Of course, for this to be done
correctly would require a calculation of the form factors, as well as
$f_{B^*}$ and $g_{B^*B\pi}$, with the same lattice parameters.

Another constraint comes from a soft pion theorem for $f_0$ which states
that
\beq
f_0(q^2_{max})=\frac{f_B}{f\pi}
\ ,\label{eq:softp}\eeq
in the chiral limit \cite{Wolfenstein:1992xh,Burdman:1992gh}. There is
some controversy as to whether the theorem is upheld by the lattice
results (see, for instance,
\cite{Hashimoto:1999bk,Bowler:1999tx,Bernard:1999ic}).  This issue
should be resolved within the coming year.

Further constraints, as we have already mentioned, come from the
scaling of heavy-to-light form factors with the mass of the heavy
quark and the energy of the final state meson (in the heavy-meson
frame), in the combined heavy-quark and large-energy limit
\cite{Charles:1998dr, Chernyak:1990ag,Ali:1994vd}. For fixed
$q^2\sim 0$, this scaling becomes a universal scaling of all form
factors with the heavy-meson mass:
\beq
F(q^2)M_B^{3/2}=A(q^2)\l(1+\frac{B(q^2)}{M_B}+\ord{\frac{1}{M_B^2}}\r)
\ ,\label{eq:lcscal}
\eeq
where $F(q^2)$ is a generic heavy-to-light form factor.  APE99
\cite{Abada:1999xd} has used the parameterisation of \eq{eq:lcscal} to
extrapolate $f_{+,0}(0)$ in heavy-quark mass from around the charm,
where the lattice yields the form factors for a range of $q^2$ that
encompasses zero, up to the $b$. This extrapolation is shown in
\fig{fig:apelcscal} and the resulting value given in \tab{tab:f0rate}.
Of course, higher-order $1/M_B$-corrections in \eq{eq:lcscal} can be
taken into account, as shown in the figure. The value reported is
consistent with the one obtained from the results of UKQCD99 as well
as with older results, as shown in \tab{tab:f0rate}.  This is very
encouraging, for it suggests that $B$-to-light form factors could be
reconstructed in the large recoil region from the lattice $D$-to-light
form factors, in a model-independent way. A couple words of warning,
however: LCSR estimates \cite{Ali:1994vd,Khodjamirian:1998vk}, as well
as the figure, suggest that pre-asymptotic corrections in
\eq{eq:lcscal} are large, even at $M_B$, and one may worry about the
convergence of the expansion in the region where lattice results are
available. Theoretical work is also needed to understand the expansion
of \eq{eq:lcscal} better.
\FIGURE{ \epsfxsize=6.5cm\epsffile{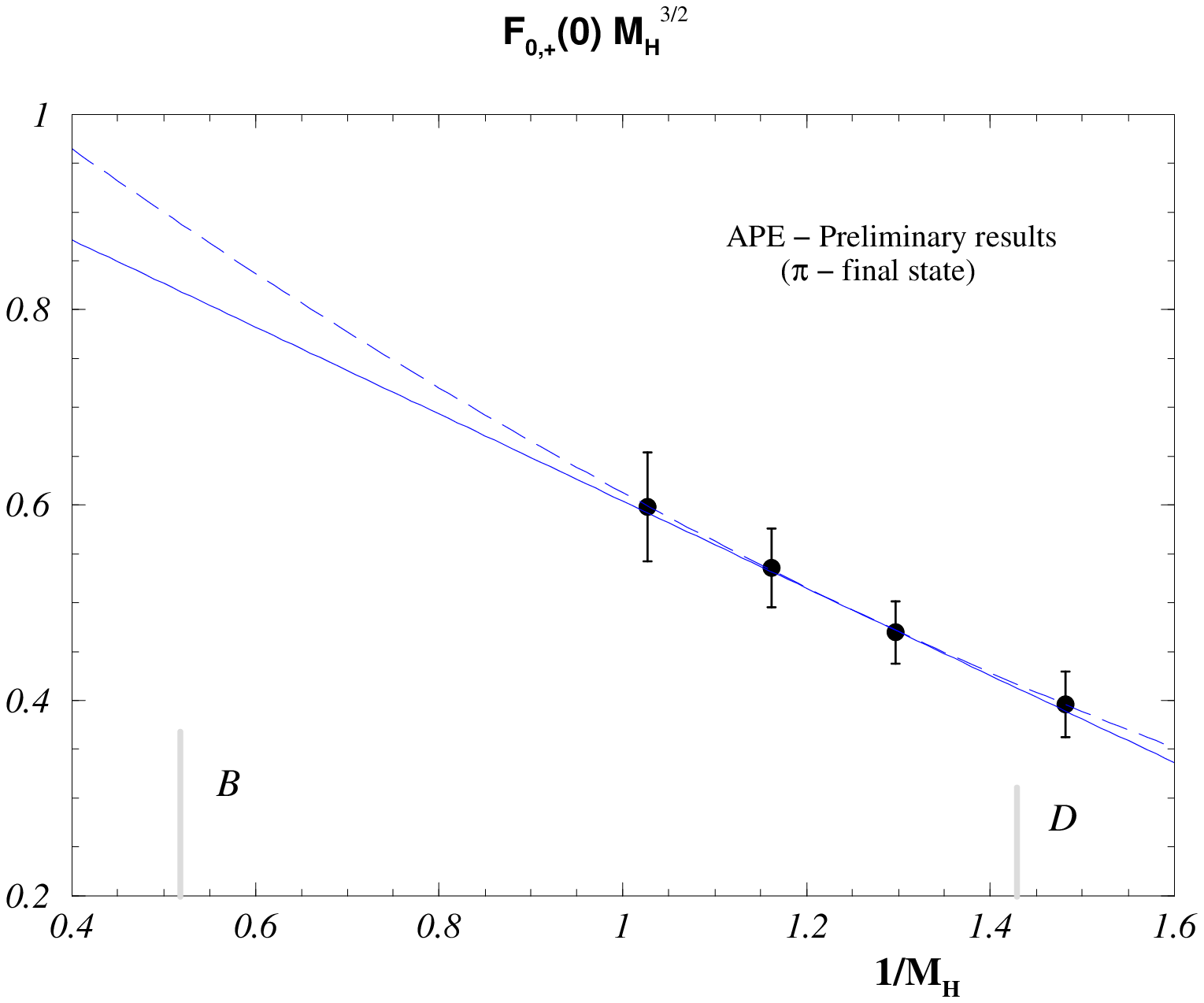}
\caption{APE99's LC scaling of $f_{+,0}(0)$ from heavy hadron mass $M_H\sim
  M_D$ to $M_B$, according to \eq{eq:lcscal} \cite{Abada:1999xd}. The
  points are the simulation results. The curves are the result of a
  linear (solid) and a quadratic (dashed) fit to \eq{eq:lcscal}.}
\label{fig:apelcscal}}

\medskip

Before closing this section, it should be mentioned that preliminary
results of a calculation for $B\to\pi$ form factors with two flavours
of dynamical quarks (partial unquenching) were presented very recently
\cite{Bernard:1999ic}.

\subsubsection{First lattice determination of the $B^*B\pi$ coupling}

Another novelty since the last Heavy Flavour conference is a first lattice
determination of the $B^*B\pi$ coupling, defined as the $q^2\to m_\pi^2$
limit of $g_{B^*B\pi}(q^2)$ in
$$
\la B^0(p)\pi^+(q)|B^{*+}(p')\ra=
$$
\beq
-g_{B^*B\pi}(q^2)q\cdot\eta\,(2\pi)^4\delta(p'-p-q)
\ .\label{eq:gbbstpidef}
\eeq
As we have already seen in \eq{eq:residue}, it is related to the residue
of $f_+$ at the $B^*$ pole. Furthermore, it determines the coupling, $g$,
of $B$, $B^*$ and $\pi$ in Heavy Meson Chiral lagrangians through
\beq
g_{B^*B\pi}(q^2\to 0)\times\frac{f_\pi}{2M_B}\stackrel{m_b\to\infty}
{\longrightarrow}\,g
\ .\label{eq:gdef}\eeq

LSZ reduction and PCAC relate the matrix element of \eq{eq:gbbstpidef}
to $\la B^0(p)|A_\mu|B^{*+}(p+q)\ra$.  The authors of
\cite{deDivitiis:1998kj} compute the latter, and do so in the static
quark limit. This exploratory
study, performed with a rather large lattice spacing, yields
\beq
g=0.42(4)(8)
\label{eq:gres}
\ ,\eeq
which compares favourably with the combined estimate of
\cite{Casalbuoni:1996pg}, $g\simeq 0.38(8)$ and is roughly consistent
with the value, $g=$ $0.27^{+4+5}_{-2-2}$, obtained from an analysis
of the experimental measurement of $D^{*(0,+)}_{(s)}\to
D^{(0,+)}_{(s)}\pi^0$ and the corresponding radiative decays
\cite{Stewart:1998ke}\footnote{The author excludes a second possible
  solution, $g=$ $0.76^{+3+2}_{-3-1}$, at the two-sigma level with the
  experimental limit, $\Gamma_{D^*+}<13\mev$.}.  It can also be
compared to the value of $g$ obtained from the direct determination of
the residue of $f_+$ given below \eq{eq:residue}.  Using \eq{eq:gdef}
at finite $m_b$ and using the value of $f_{B^*}$ given after
\eq{eq:residue}, I find $g=0.37^{+7+34}_{-7-7}$, consistent within 
statistical errors.

JLQCD are undertaking an indirect determination of $g$ as part of
their NRQCD study of $B\to\pi\ell\nu$ decays, from the residue of the
$B^*$ pole \cite{Hashimoto:1999bk}. There are also determinations of
$g$ from the residue of the $B^*$ pole from older lattice results in
\cite{Lellouch:1996yv}.

\subsubsection{$B\to\rho\ell\nu$ and $B\to K^*\gamma$}

Although no new lattice results for these decays have been obtained
since the last Heavy Flavour conference, it is worth reviewing a few
results. The definitions of the form factors discussed below can be
found, for instance, in \cite{Flynn:1996dc}.

UKQCD \cite{Flynn:1996dc} proposed obtaining $|V_{ub}|$
from a comparison of lattice and experimental results for the
differential rate for $B^0\to\rho^-\ell^+\nu$ decays for $q^2\gsim
12\gev^2$. To that effect, they parameterised this rate as
\beq
\frac{d\Gamma}{dq^2}=\frac{G_F^2|V_{ub}|^2}{192\pi^3 M_B^3}q^2
\sqrt{\lambda(q^2)}a^2(1+b(q^2-q^2_{max}))
\eeq
where $\lambda(q^2)=(M_B^2+M_\rho^2-q^2)^2-4M_B^2M_\rho^2$ and 
where the parameters $a$ and $b$ are determined
by fitting to lattice results obtained for $q^2\gsim 14\gev^2$. They find
\bea
a&=&4.6^{+0.4}_{-0.3}\pm 0.6\gev\\
b&=&(-8^{+4}_{-6})\times 10^{-2}\gev^{-2}
\,\nn \eea
where $a$ includes a 6.5\% systematic uncertainty associated with
light spectator mass dependence and a 10\% uncertainty associated with
discretisation effects. $a$, here, plays the role of ${\mathcal
  F}_{D^{*}}(1)$ in the determination of $|V_{cb}|$ from $B\to
D^*\ell\nu$ decays, and $b$ the role of ${\mathcal F}_{D^{*}}$'s slope
at $w=1$.

For $\rho=K^*$, heavy-quark spin symmetry predicts that the form factors
$V$ and $A_1$ for $B\to\rho\ell\nu$ and $T_{1,2}$ for $B\to
K^*\gamma$ are related as \cite{Isgur:1990ed}
\bea
\frac{V}{2T_1}(w)&=&1+\ord{\frac{1}{m_b}}\nn\\
\frac{A_1}{2iT_2}(w)&=&1+\ord{\frac{1}{m_b}}
\label{eq:hqrats}\ ,\eea
for $w=v_B\cdot v_{\rho,K^*}\sim 1$. Results for these ratios are
shown in \fig{fig:vovert1}, for different values of the initial
heavy-meson mass and for a $\rho$ and $K^*$ composed of quarks
slightly more massive than the strange. Both relations are well
satisfied in the infinite heavy-quark-mass limit.  The surprising
result is that the ratio $A_1/2iT_2$ still satisfies the heavy-quark
prediction, even
at the $D$ mass.
\FIGURE{\vbox{\offinterlineskip
\hbox to0.8\hsize{\hfill
\epsfysize=0.70\hsize
\epsffile[35 35 232 510]{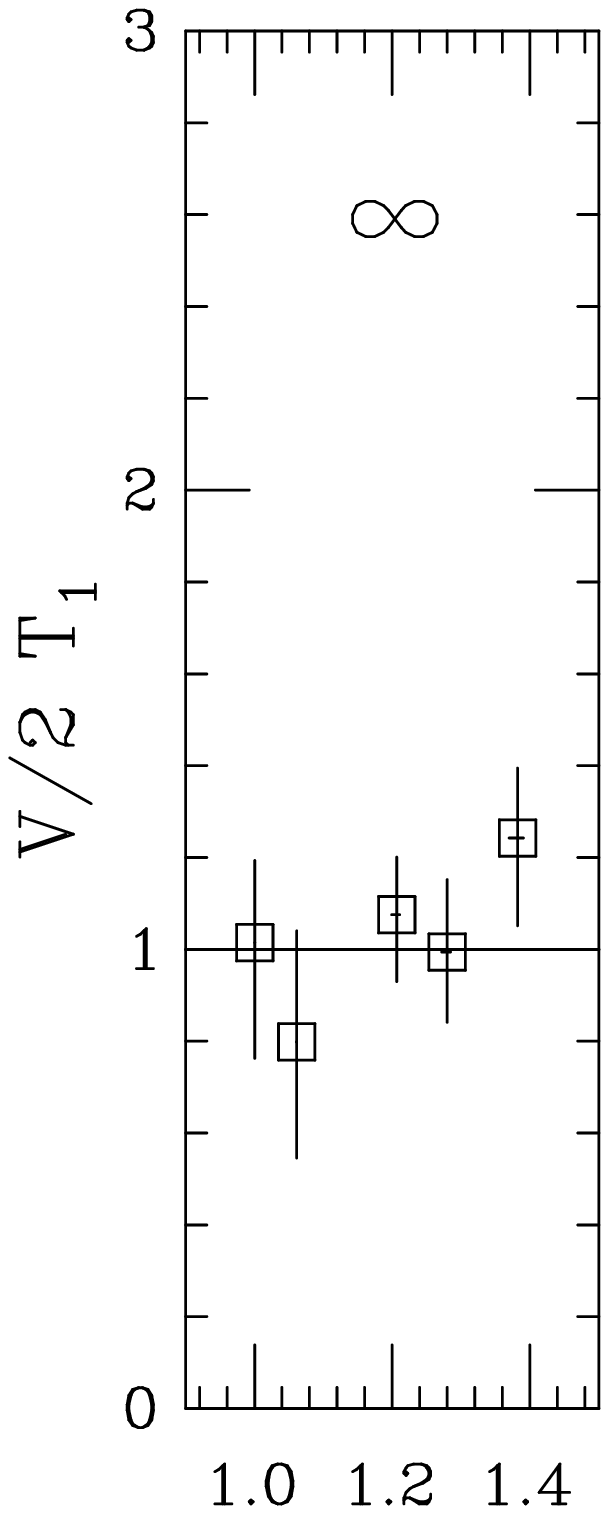}
\epsfysize=0.70\hsize
\epsffile[73 35 232 510]{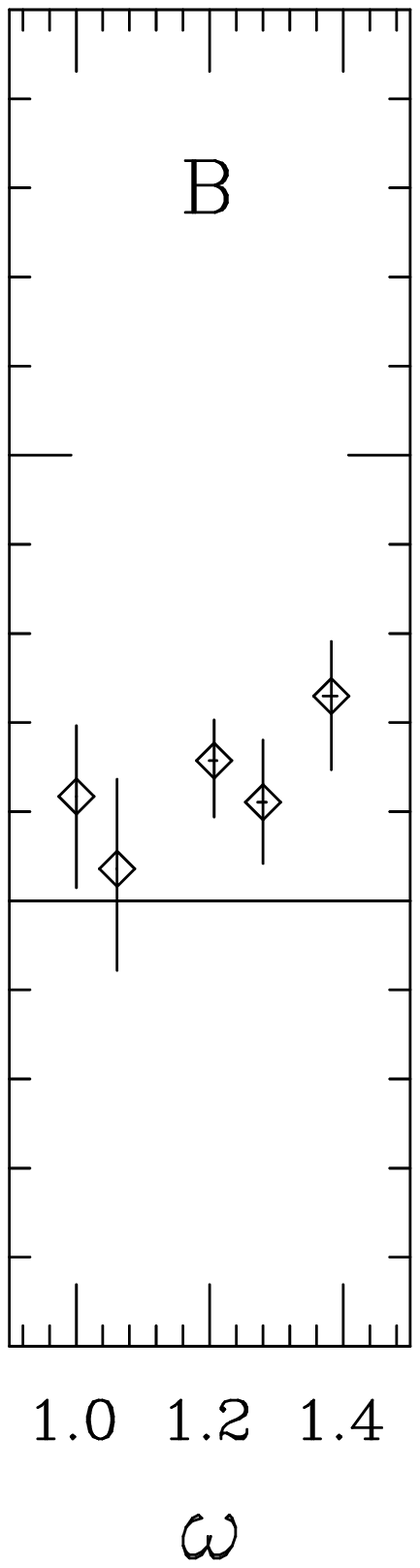}
\epsfysize=0.70\hsize
\epsffile[73 35 232 510]{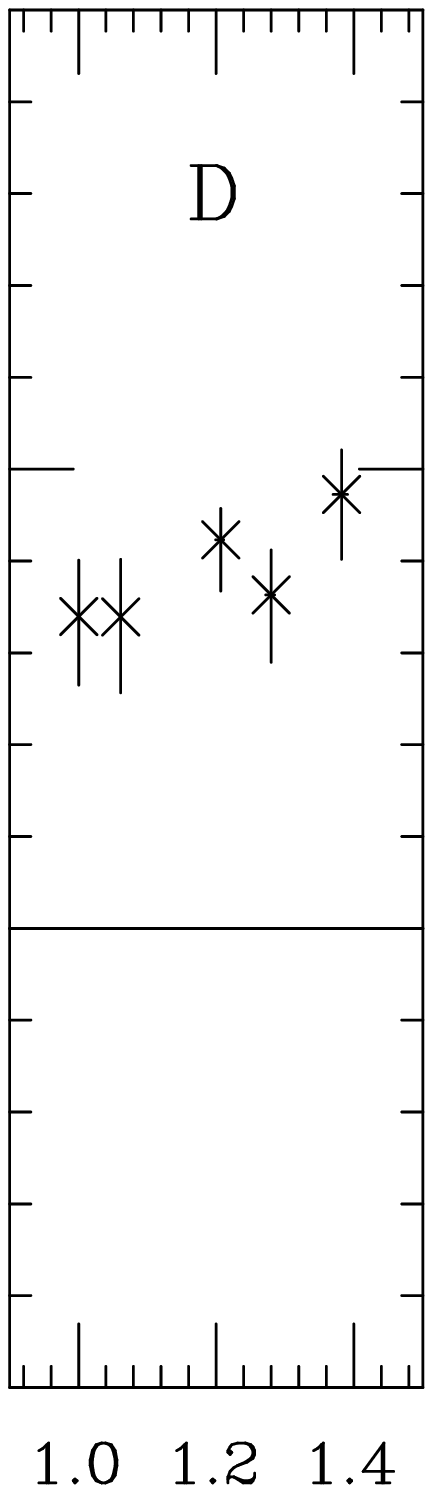}
\hfill
}\kern1em
\hbox to0.8\hsize{\hfill
\epsfysize=0.70\hsize
\epsffile[35 35 232 510]{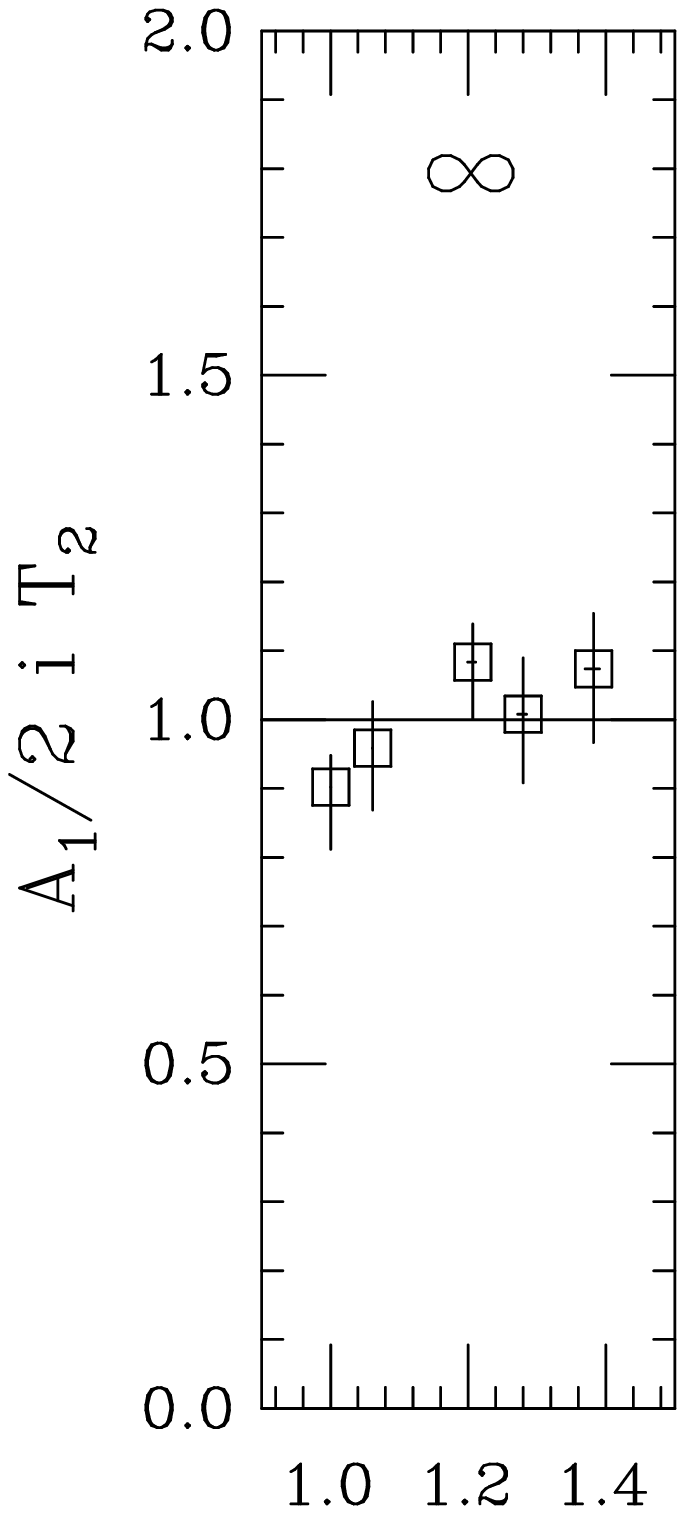}
\epsfysize=0.70\hsize
\epsffile[73 35 232 510]{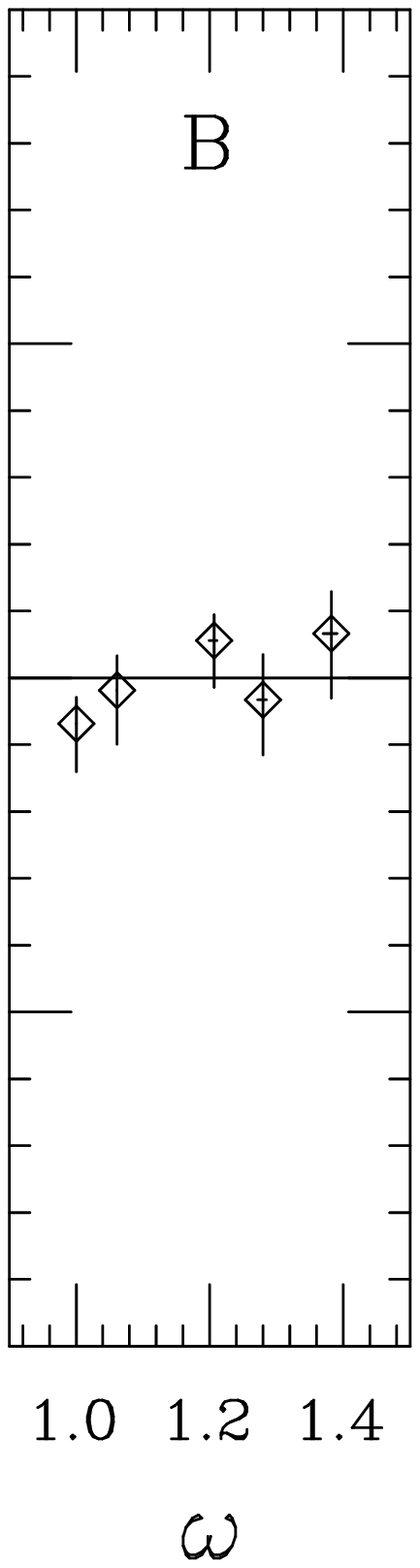}
\epsfysize=0.70\hsize
\epsffile[73 35 232 510]{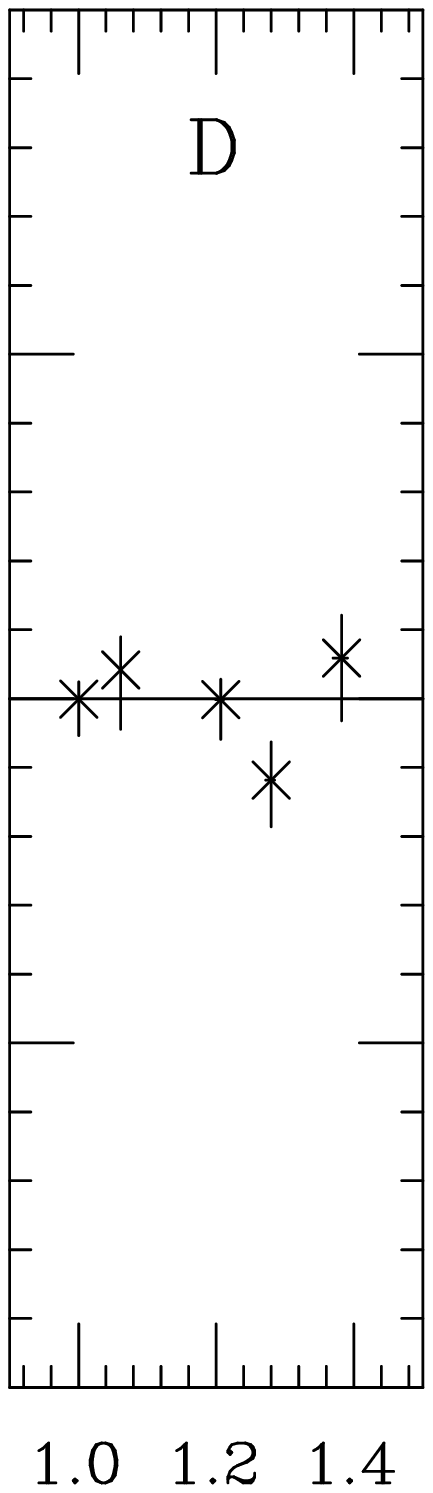}
\hfill
}}
\caption{Ratios $V/2T_1$ and $A_1/2iT_2$ for five values of $w$ and
  three values of the initial heavy-meson mass
  \cite{Flynn:1996dc}.  The horizontal lines are the
  heavy-quark limit values.}
\label{fig:vovert1}
}
One can further include all $1/m_b$ corrections from kinematics
\cite{Isgur:1990ed,Burdman:1992hg,Griffin:1994kj} in the ratios of
\eq{eq:hqrats}. With these corrections the relation between $V$ and
$T_1$ is improved by a factor of about two while the relation between
$A_1$ and $T_2$ is not significantly changed. Similar relations,
including the $1/m_b$ corrections from kinematics, have been
considered with LCSR, but for large recoils ($1.7\lsim w \lsim 3.5$),
and are found to hold to good accuracy \cite{Ali:1994vd,Ball:1998kk}.

As already discussed for $B\to\pi\ell\nu$ decays, it may be of
interest to abandon model-independence in favour of a simple
description of the relevant form factors over the full kinematical
range.  UKQCD \cite{DelDebbio:1997kr} thus considered combined fits of
lattice results for the form factors relevant for $B\to\rho\ell\nu$
and $B\to K^*\gamma$ decays, obtained at high $q^2$. The
parameterisations used are consistent with heavy-quark scaling at
small recoils (\eq{eq:hqscal}) and light-cone scaling at large recoils
(\eq{eq:lcscal}), as well as with the kinematical constraint at
$q^2=0$, $T_1(0)=iT_2(0)$\footnote{They also consider $B\to\pi\ell\nu$
  decays, but the fits discussed around \eq{eq:bpiparam} are performed
  with improved lattice results.}.  For degenerate $\rho$ and $K^*$,
leading order HQET predicts that the seven form factors required to
parameterise the relevant matrix elements are related to four
Isgur-Wise functions \cite{Isgur:1990ed,Griffin:1994kj}. Keeping only
one of these functions, which is fixed by parameterising $A_1(q^2)$ by
a pole form with free pole position and residue, the authors of
\cite{DelDebbio:1997kr} obtain a two-parameter description of the
seven form factors. The results of the fit to this parameterisation
are shown in \fig{fig:latconpar} for a final state
$\rho$\footnote{$A_2(q^2)$ and $T_3(q^2)$ are not included in the fit.
  $T_3(q^2)$ has not yet been calculated.}. Also shown are the lattice
points of APE \cite{Allton:1995ui} and ELC \cite{Abada:1994dh} as well
as the LCSR results of \cite{Ball:1998kk}.  Agreement with the lattice
results in the region of overlap is encouraging as is the agreement at
smaller $q^2$, with the lattice-constrained parameterisation.  It is
worth pointing out that the error bars on the UKQCD results are
statistical only. As discussed in \cite{Flynn:1996dc}, one may assign
a $10\%$ uncertainty for discretisation errors and a 5-20\%
uncertainty, depending on the form factor, associated with light
spectator mass dependence. We also emphasise that the extrapolated
form factors are no longer model independent, though models
related to the one used have recently received theoretical support
from the formalism developed in \cite{Charles:1998dr}.
\FIGURE{
\vbox{\offinterlineskip
\hbox to 2\hsize{\hfill
\epsfysize=0.65\hsize
\epsffile[100 210 530 620]{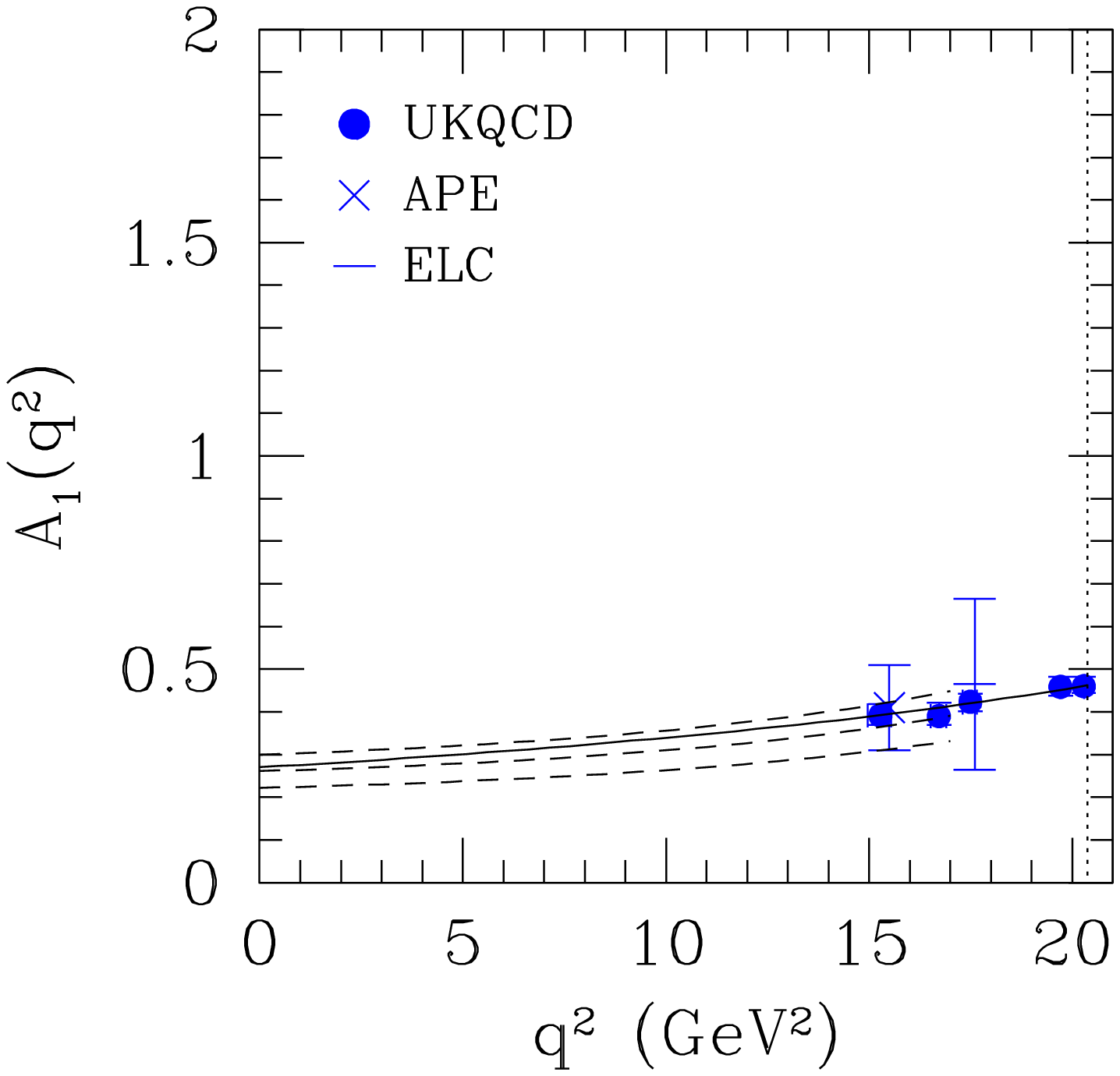}
\epsfysize=0.65\hsize
\epsffile[100 210 530 620]{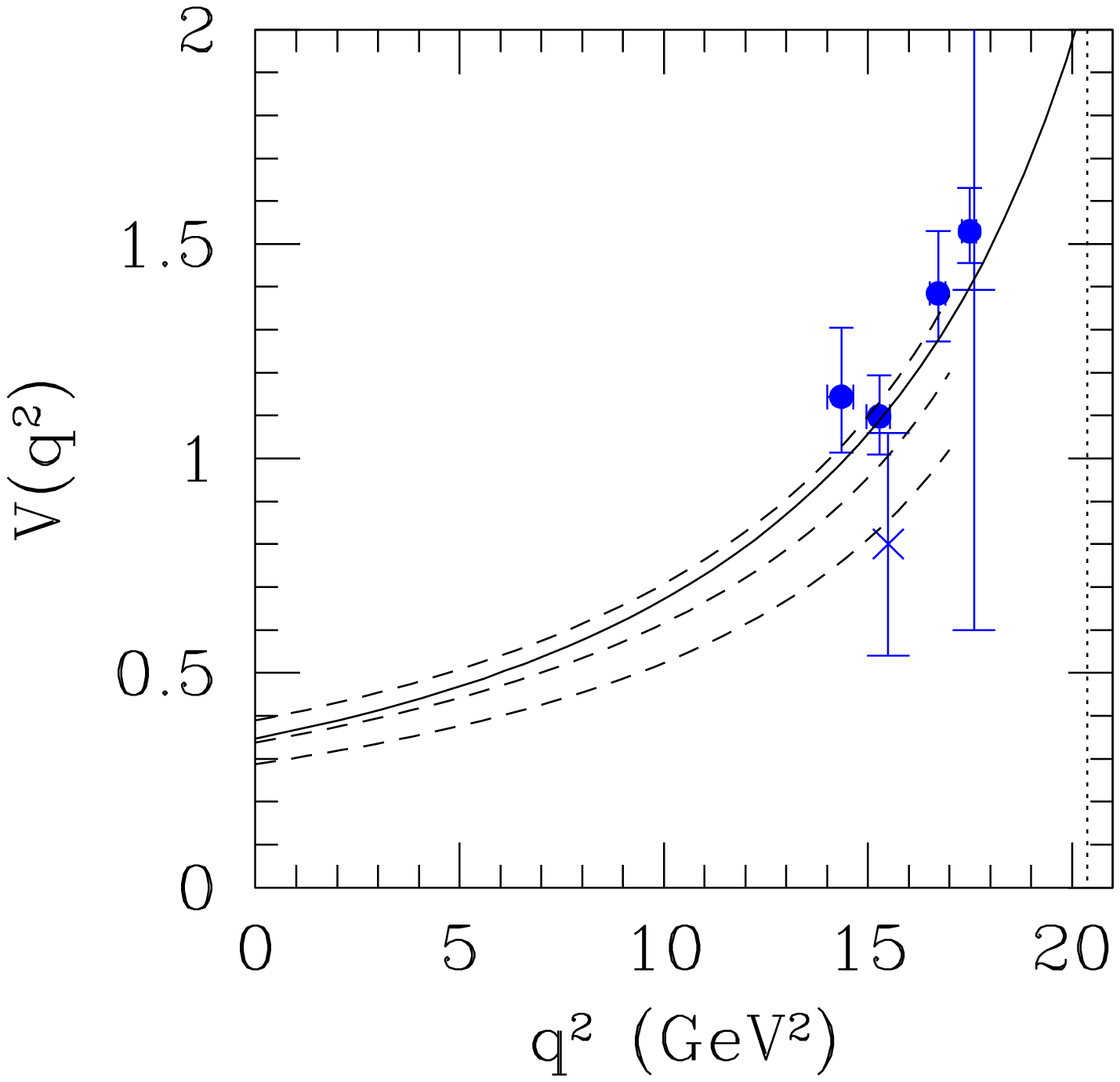}
\epsfysize=0.65\hsize
\epsffile[100 210 530 620]{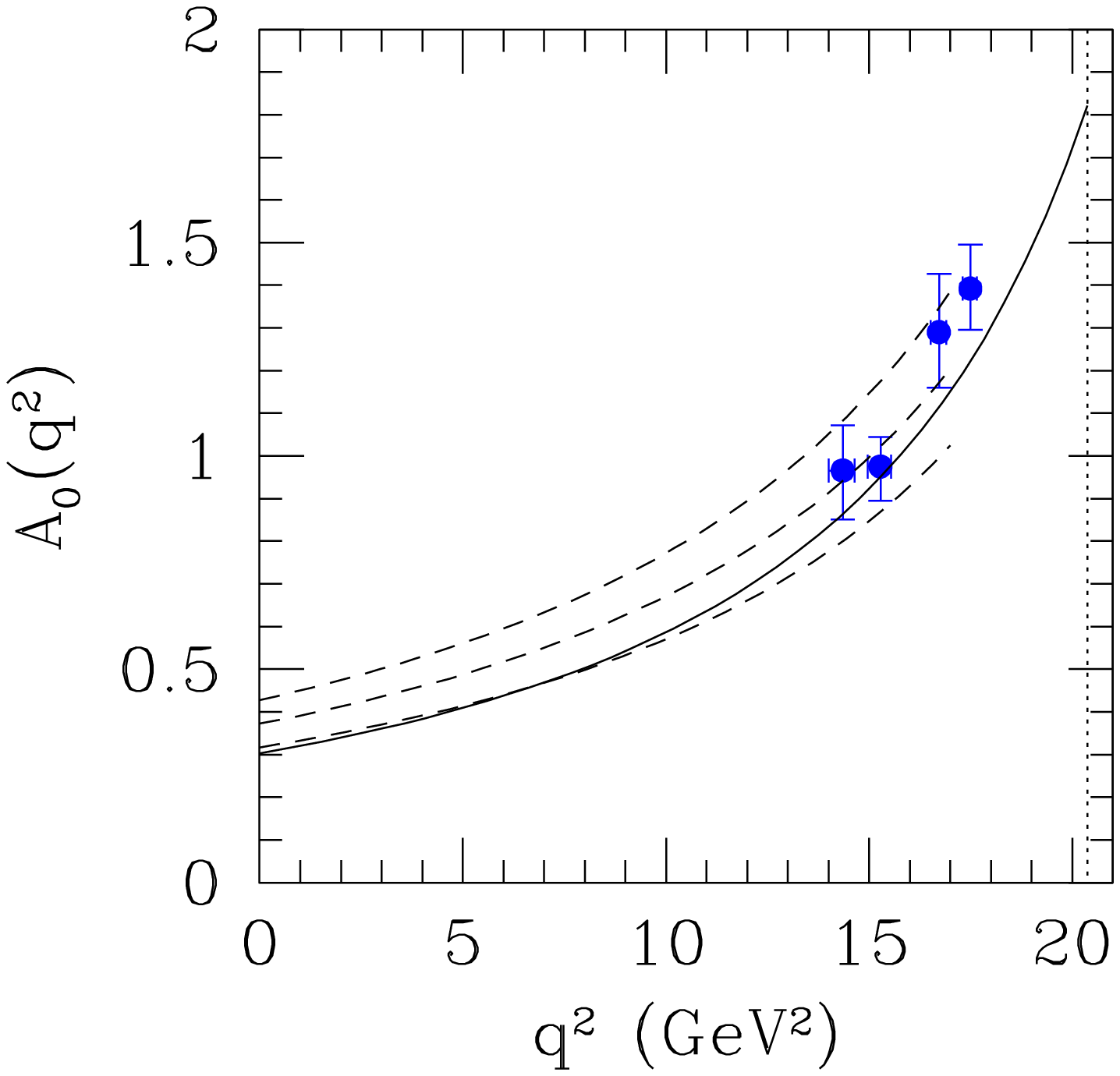}\hfill
}\kern1em
\hbox to 2\hsize{\hfill
\epsfysize=0.65\hsize
\epsffile[100 210 530 620]{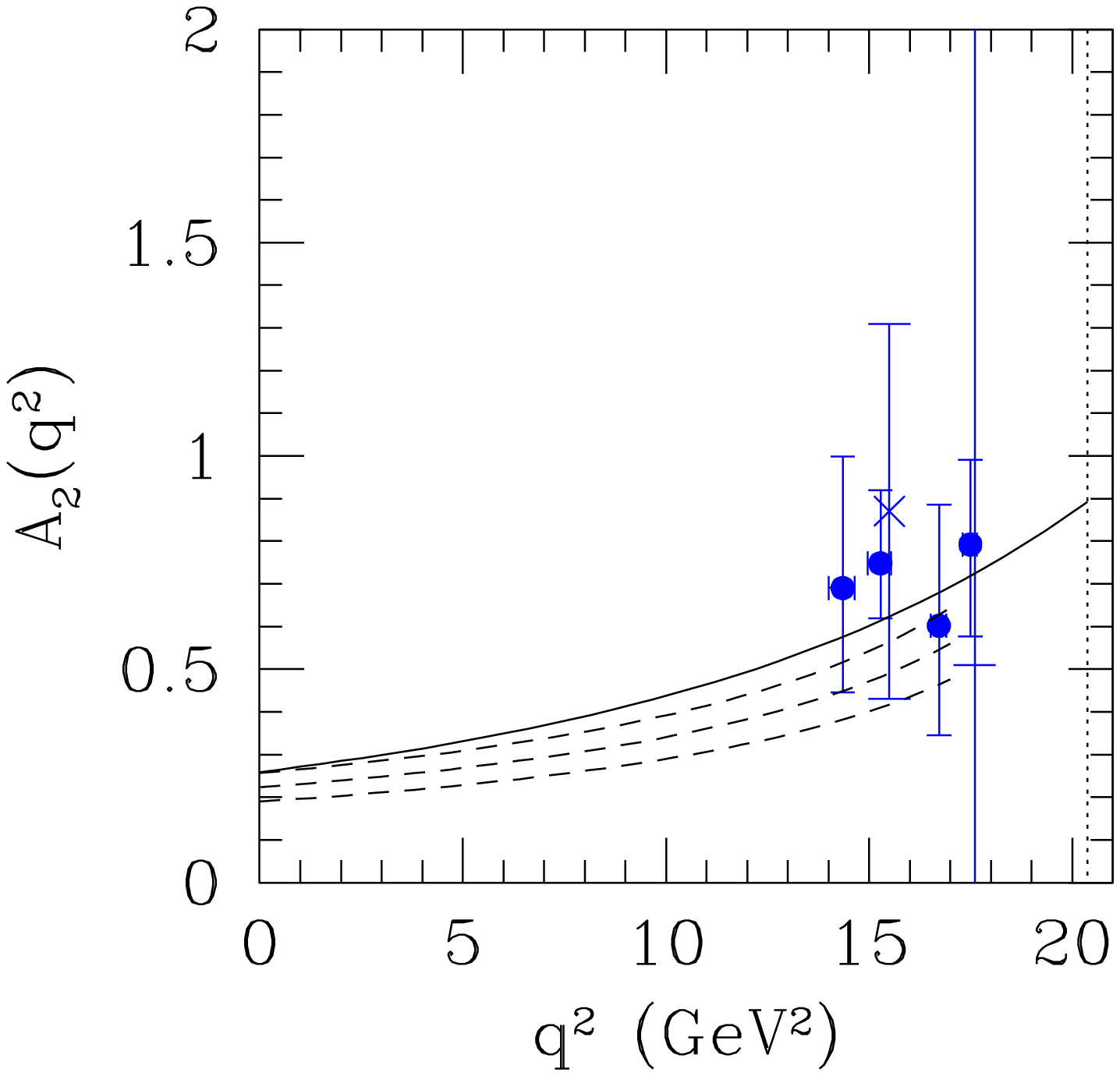}
\epsfysize=0.65\hsize
\epsffile[100 210 530 620]{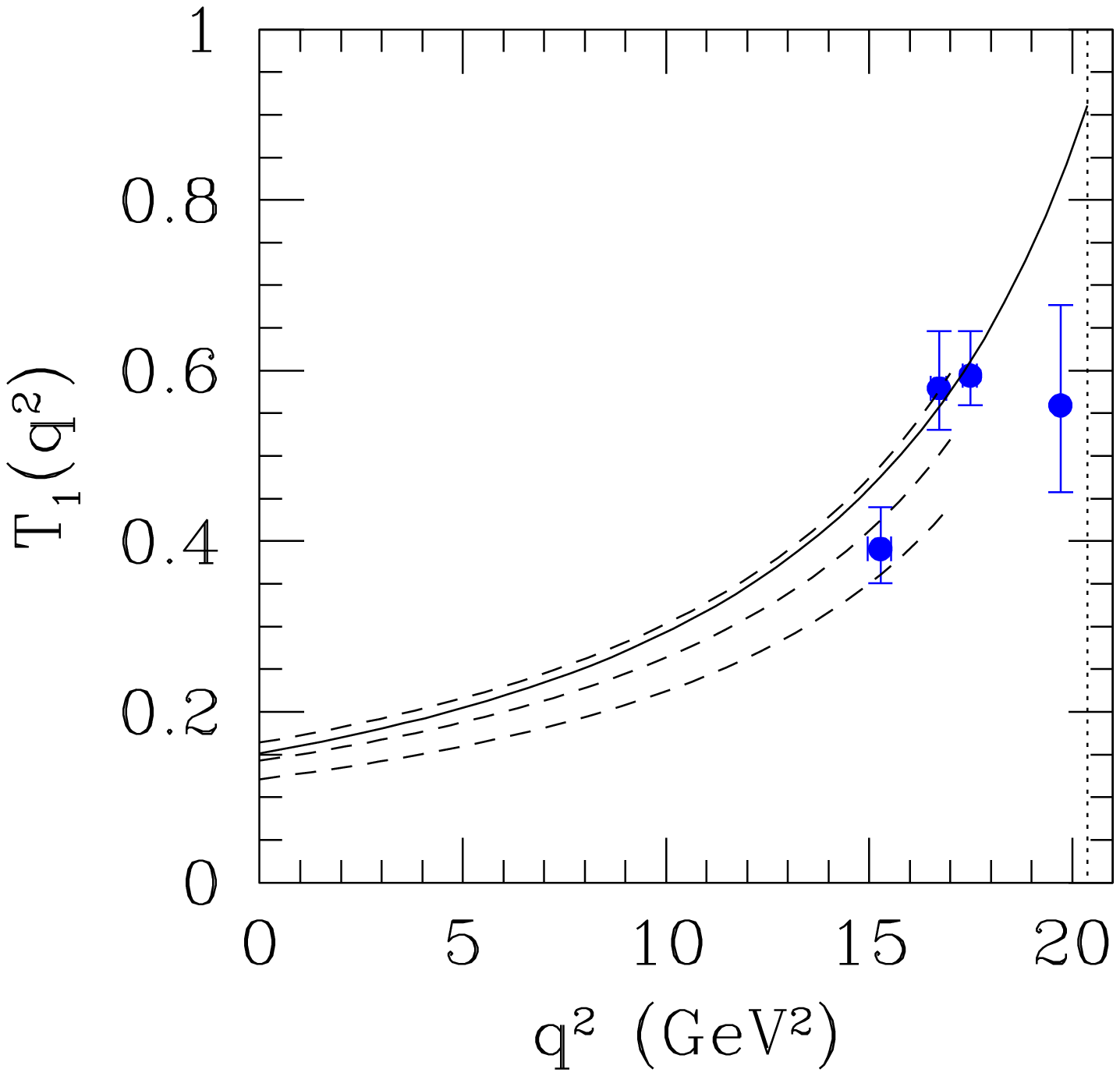}
\epsfysize=0.65\hsize
\epsffile[100 210 530 620]{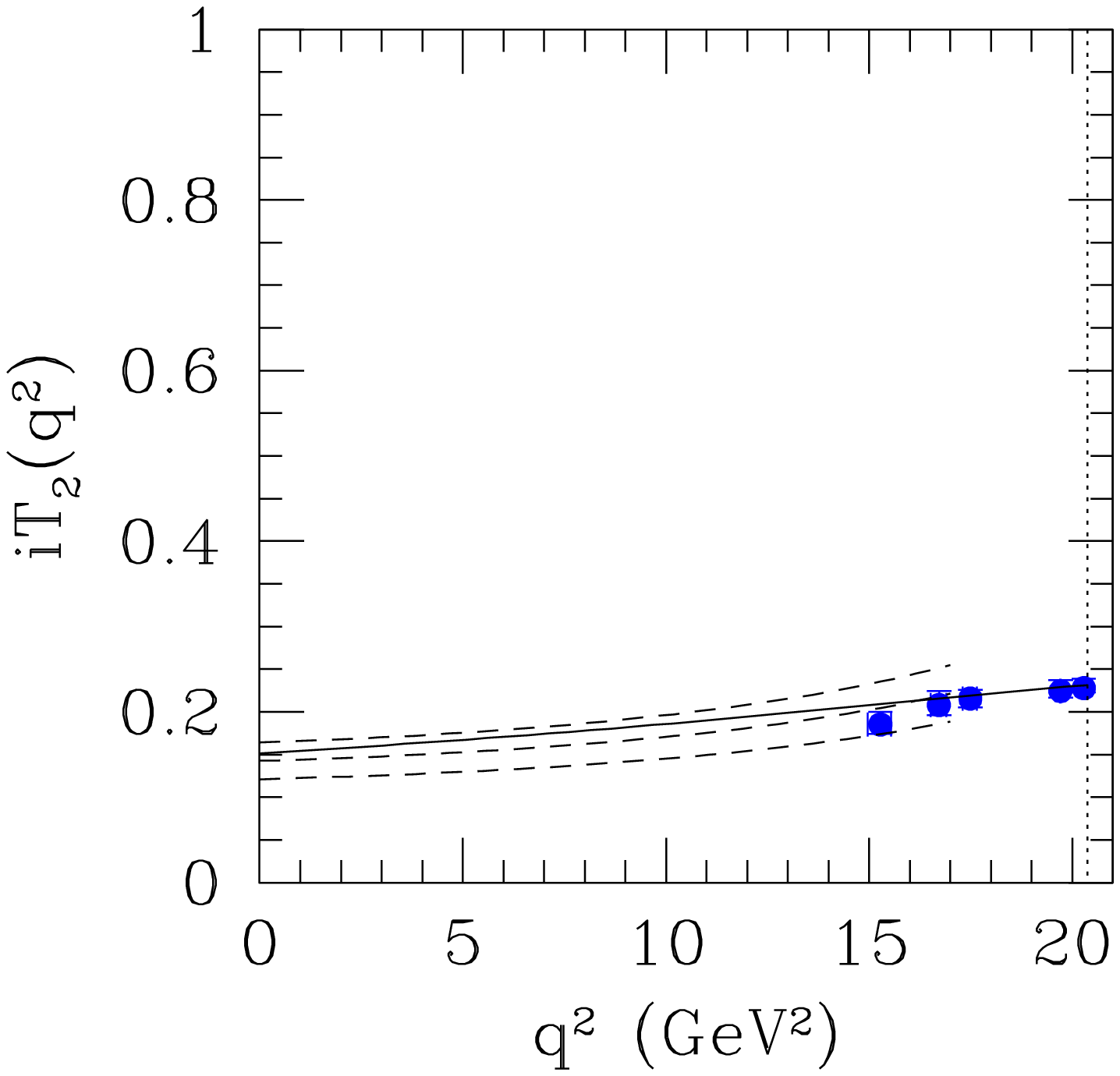}
\hfill
}}
\caption{$B\to\rho$ form factors. The solid curves
  are fits of the UKQCD results to the constrained parameterisation
  described in the text \cite{DelDebbio:1997kr}. The dashed curves are
  the LCSR results of \cite{Ball:1998kk} with a 15\% error band.}
\label{fig:latconpar}
}

Similar results are obtained for a final state $K^*$ and agreement
with LC sumrules is also good. These results enable a determination of
$T_1(0)$ which determines the rate for $B\to K^*\gamma$. In
\tab{tab:bksg} we compare $T_1(0)$ from the combined fit of
\cite{DelDebbio:1997kr} to older results obtained with less constrained
extrapolations. The UKQCD value yields $R_{K^*}=$ $\Gamma(B\to
K^*\gamma)/$ $\Gamma(b\to s\gamma)=16^{+3}_{-4}\%$, to leading order
in $\alpha_s$ and up to $\ord{1/m_B^2}$ corrections
\cite{Ciuchini:1994xa}. Errors are statistical only. This result is
consistent with the experimental result from CLEO \cite{cleorkst},
$R_{K^*}=13(4)\%$.
\TABLE{\begin{tabular}{rcll}
\hline
& $T_1(0)$ & $T_2(q^2_{max})$\\
\hline
UKQCD98 \cite{DelDebbio:1997kr}& $0.16^{+2}_{-1}$ & 0.25(2)\\
LANL96 \cite{Gupta:1995zd} & 0.09(1) & \\
APE96 \cite{Abada:1996fa} & 0.09(1)(1) & \\
BHS94 \cite{Bernard:1994yt} & 0.101(10)(28) & 0.325(33)(65)\\
\hline
\end{tabular}
\caption{$T_2(q^2_{max})$ is obtained using heavy-quark scaling from
  around the charm region to the $b$. For $T_1(0)=iT_2(0)$, only
  results from extrapolations consistent with LC scaling
  (\eq{eq:lcscal}) are given.}
\label{tab:bksg}
}

\subsubsection{Other decays}

APE \cite{Abada:1999xd} is also considering the penguin induced matrix
elements for $B\to K(\pi)\ell^+\ell^-$ decays.

\subsection{Heavy-to-heavy decays}

The recoils involved in semileptonic heavy-to-hea\-vy quark decays are
much smaller than those in heavy-to-light decays: the lattice can
cover the full kinematical range. Heavy quark symmetry is also much
stronger, here, in that it normalises the relevant matrix elements at zero
recoil and applies to the full kinematical range. Nevertheless, the
lattice can provide interesting information on the behaviour of these
matrix elements away from zero recoil as well as provide tests of
heavy quark symmetry. Attempts are also being made to quantify
deviations from heavy-quark normalisations for physical quark masses
at the zero-recoil point.

\subsubsection{$B\to D^{(*)}\ell\nu$: form factors at zero recoil}

To extract $|V_{cb}|$ from $B\to D^{(*)}\ell\nu$ decays, one
traditionally extrapolates the differential decay rates to the
zero-recoil point, $w=v_B\cdot v_{D^{(*)}}=1$, obtaining
$|V_{cb}|\mathcal{F}_{D^{(*)}}(w=1)$. One then uses the fact that the
form factors $\mathcal{F}_{D^{(*)}}(1)$ are equal to 1 in the heavy
quark limit to obtain $|V_{cb}|$. A precise determination of
$|V_{cb}|$, however, requires calculation of the corrections to
this limit. Apart from calculable, perturbative corrections, there are
non-perturbative corrections proportional to inverse powers of
the heavy quark masses, $m_{c,b}$. Uncertainties on these power corrections
currently limit a more precise determination of $|V_{cb}|$.

It was proposed recently that these corrections be obtained by
studying the heavy-quark-mass dependence of the relevant form factors
on the lattice \cite{Hashimoto:1999yp}, instead of evaluating, for
instance, the contributions of subleading HQET operators in the
$1/m_{c,b}$ expansion. In \cite{Hashimoto:1999yp}, this idea is
applied to the determination of power corrections to
$\mathcal{F}_{D}(1)$. In \cite{Simone:1999nv}, the same authors 
present preliminary results for the corrections to
$\mathcal{F}_{D^*}(1)$. Since their method requires measuring small
deviations from 1, excellent control of both statistical and
systematic errors is necessary. Thus, they suggest studying the mass
dependence of double ratios of three-point functions which at
asymptotic times reduce to a double ratio of zero-recoil matrix
elements\footnote{To determine zero-recoil, power corrections in the
  form factor $h_-(w)$, relevant for $B\to D\ell\nu$ decays,
they use slightly different ratios.}:
\beq R_{J_\mu}^{B^{(*)}\to D^{(*)}}\longrightarrow \frac{\la
  D^{(*)}|J_\mu^{cb}|B^{(*)}\ra \la B^{(*)}|J_\mu^{bc}|D^{(*)}\ra}{\la
  B^{(*)}|J_\mu^{bb}|B^{(*)}\ra \la D^{(*)}|J_\mu^{cc}|D^{(*)}\ra} \ 
,\label{eq:dblerat}
\eeq
where $J_\mu^{qq'}=\bar q\gamma_\mu q'$ or $\bar q\gamma_\mu\gamma_5
q'$. 

For instance, $R_{V_0}^{B\to D}$ yields $|h_+(1)|^2$ up to a
multiplicative renormalisation, where $h_+(w)$ is the usual $B\to
D\ell\nu$, heavy-quark form factor. The dependence of the square root
of this ratio on $1/m_c$ is shown in \fig{fig:hphqmdep}.  Statistical
errors are indeed very small, and the deviation from 1 is significant
statistically, which is certainly very encouraging.  However, the
tuning of the action and operators and the analysis of systematics in
the hybrid approach is complex, and the results of
\cite{Hashimoto:1999yp,Simone:1999nv} should be confirmed by other groups.
\FIGURE{\epsfxsize=6.5cm\epsffile{h+_heavy_mass_dependence.epsi}
\caption{$R_{V_0}^{B\to D}$ of \eq{eq:dblerat} as a function of $1/am_c$
  for an initial heavy quark with mass $am_b=2.11$
  \cite{Hashimoto:1999yp}.  The light quark's mass is that of the
  strange.}
\label{fig:hphqmdep}
}

\subsubsection{$B\to D^{(*)}\ell\nu$: recoil dependence}

While the shapes of $B\to D^{(*)}\ell\nu$ form factors obtained from
the lattice are not required for determining $|V_{cb}|$ from these
decays\footnote{Extrapolation of the experimental data for the
  differential decay rate to the zero-recoil point with the help of a
  dispersive parameterisation such as the ones presented in
  \sec{sec:dispbd} is adequate.}, it is informative to compare these
shapes to those determined by experiment. Such a comparison enables
one to verify the reliability of the extrapolation of experimental
data to the zero-recoil point as well as to check the lattice method.

UKQCD \cite{Douglas:1999wr} has presented preliminary results of a
non-perturbatively $\ord{a}$-improved calculation of a 
candidate Isgur-Wise function relevant for $B\to D^{(*)}\ell\nu$ decays:
\beq 
\mbox{``}\xi(w)\mbox{''}
\equiv\frac{(1+\beta_+(1))}{(1+\beta_+(w))}\frac{h_+^{lat}(w)}{h_+^{lat}(1)}
\ ,\label{eq:iwdef}
\eeq
where $\beta_+(w)$ are the radiative corrections which match the QCD
vector current, $\bar c\gamma^\mu b$, to its HQET counterpart and
where $h_+^{lat}$ is the heavy-quark form factor $h_+$ up to a
multiplicative renormalisation. The ratio of \eq{eq:iwdef} enforces
non-perturbative lattice to continuum matching and should lead to
cancellations of discretisation errors as well as subtract some
zero-recoil power corrections to the heavy quark limit.  Following
\cite{Booth:1994zb,Bowler:1995bp}, they study the heavy-quark mass
dependence $\mbox{``}\xi(w)\mbox{''}$ for heavy-quark masses around
the charm and find that it is consistent with 0, in the range $1\le
w\le 1.2$. Thus, $\mbox{``}\xi(w)\mbox{''}$ is to a good approximation
the Isgur-Wise function $\xi(w)$, at least in the above range of $w$,
confirming the earlier results of \cite{Booth:1994zb,Bowler:1995bp}.
The calculation is performed at two values of the lattice spacing,
$\beta=6.2$ (finer) and 6.0 (coarser).  Results for the Isgur-Wise
function relevant for semileptonic $B\to D^{(*)}$ decays are shown in
\fig{fig:garyiw}. Dependence on lattice spacing is small suggesting
that discretisation errors are small. These results are compatible,
yet more accurate than previous
determinations (see \cite{Flynn:1997ca}).
\FIGURE{
\hbox{\hfill\vbox{\epsfxsize=5.5cm\epsffile[30 30 530 530]{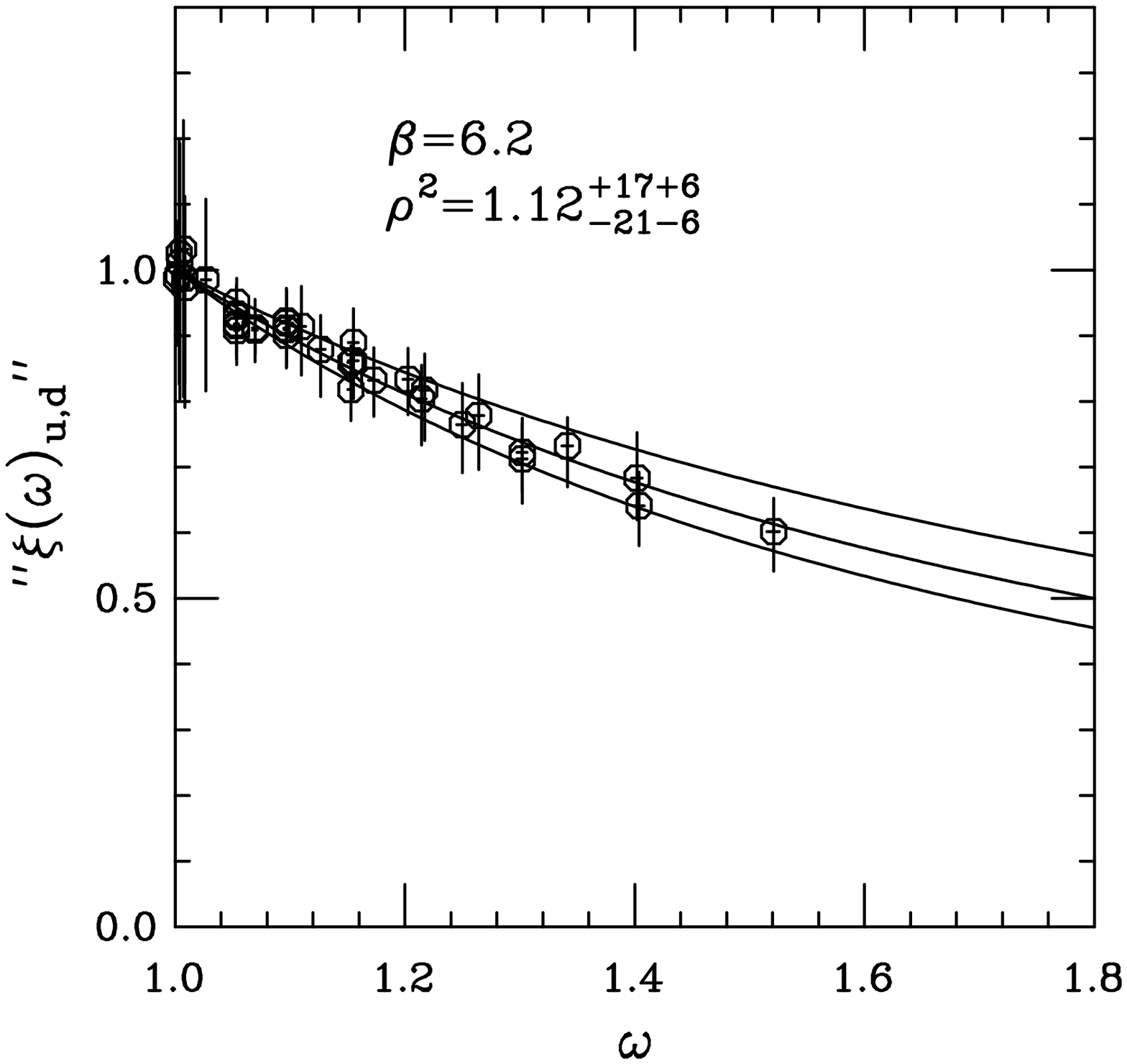}}
\vbox{\epsfxsize=5.5cm
\epsffile[30 30 530 530]{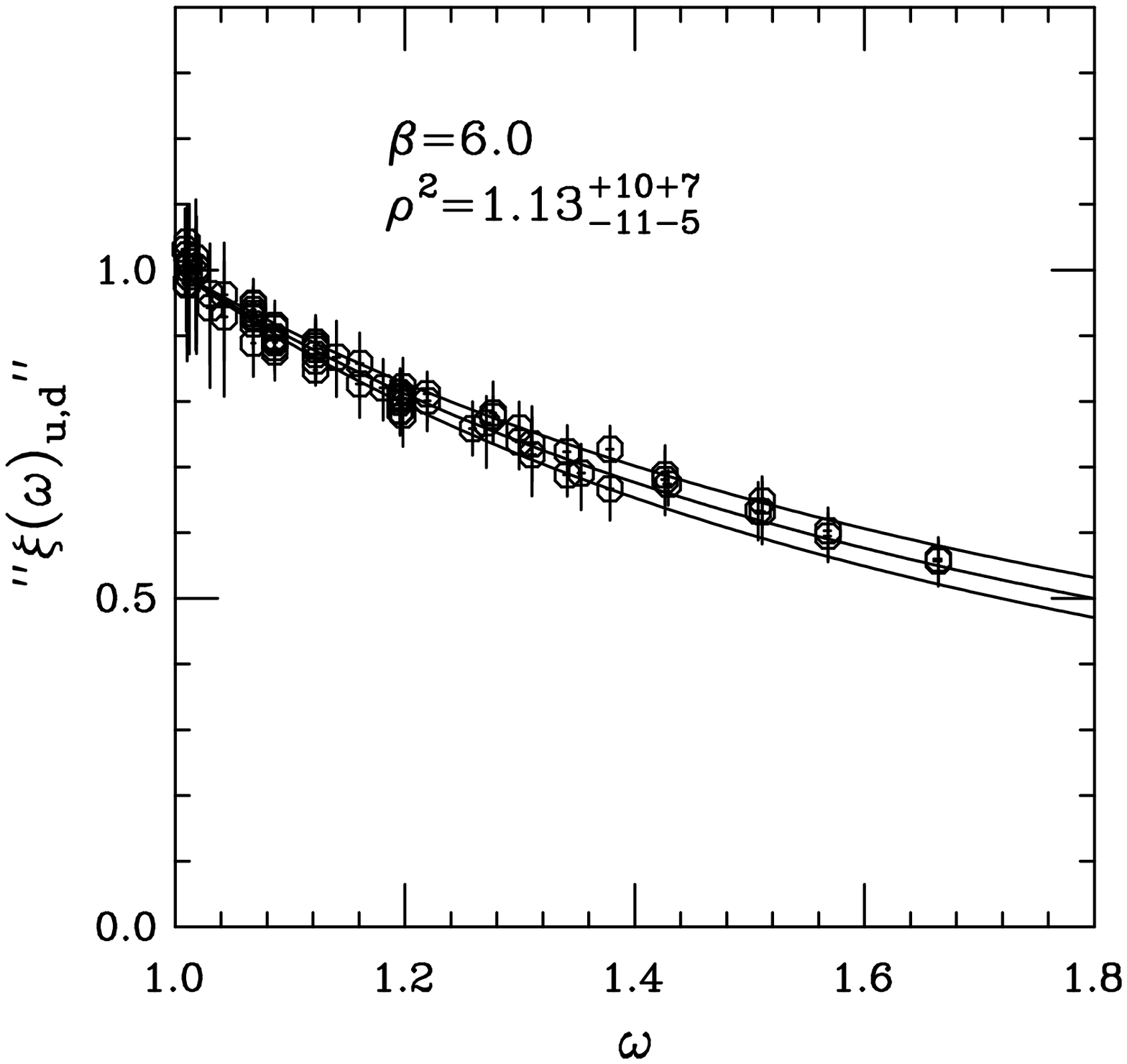}}\hfill}
\caption{Candidate Isgur-Wise function relevant for 
  $B\to D^{(*)}\ell\nu$ decays at two values of the lattice spacing
  (finer to the left) \cite{Douglas:1999wr}. The curves, with
  statistical error bands, and the values of the zero-recoil slope,
  $\rho^2$, are the result of fits to $\xi(w)=$ $\frac{2}{w+1}$
  $\mathrm{exp}\l[(2\rho^2-1)\frac{1-w}{1+w}\r]$. The second error
  bar on $\rho^2$ is an estimate of discretisation errors as in
  \cite{Bowler:1995bp}.}
\label{fig:garyiw}
}

GOK are pursuing an NRQCD study of heavy-to-heavy decays
\cite{Hein:1999se}. In addition to $B\to D$, they are also considering
decays of the $B$ to radially excited $D$ mesons.

\subsubsection{Other decays}

UKQCD has investigated $\Lambda_b\to\Lambda_c\ell\nu$ and
$\Xi_b\to\Xi_c\ell\nu$ decays and determined the relevant 
Isgur-Wise functions \cite{Bowler:1998ej}.

\section{Dispersive bounds}

The second part of my talk is concerned with the constraints
obtained on weak matrix elements from the polarisation function
\beq
\Pi^{\mu\nu}_J(q^2) = i\int d^4x\ e^{iq\cdot x}\la 0|T\l\{J^{\mu}(x)
J^{\nu\dagger}(0)\r\}|0\ra
\ ,\label{eq:2pt}
\eeq
where $J^\mu$ is chosen to be the operator which mediates the weak
transition under consideration\footnote{While the subject of inclusive
  heavy-quark sumrules is a fascinating one, it would take us too far
  afar.}.

I will further concentrate on two applications:
\begin{itemize}
\item model-independent constraints on $B\to $ $D^{(*)}\ell\nu$ form
  factors used to eliminate the uncertainty in the extrapolation of
  experimental data to $w=1$ in the extraction of $|V_{cb}|$.
\item model-independent extrapolations of heavy-to-light lattice
  results obtained in a limited $q^2$ range.
\end{itemize}

\subsection{$B\to D^{(*)}\ell\nu$}
\label{sec:dispbd}

I will briefly summarise the methodology of dispersive
constraints in the context of $B \to D\ell\nu$, with
$J^\mu=V^\mu=\bar c\gamma^\mu b$. One first decomposes the
polarisation function of \eq{eq:2pt} according to helicity as
\beq
\Pi^{\mu\nu}_V(q^2) =
(q^\mu q^\nu-g^{\mu\nu} q^2) \Pi_{1^-}(q^2) + q^\mu q^\nu \Pi_{0^+}(q^2)
\ ,\label{eq:spindecom}
\eeq
paralleling the decomposition of the matrix element $\la D|
V^\mu| B\ra$ in terms of the form factors $f_+$ and $f_0$. To
constrain $f_0$, one writes down a (once subtracted) dispersion
relation for $\Pi_{0^+}$:
$$
\chi_{0^+}(q^2) = \frac{\partial}{\partial q^2}
\l(q^2 \Pi_{0^+}(q^2)\r)
$$
\beq
=\frac{1}{\pi}\int_0^\infty dt\ \frac{t\,\mbox{Im}\Pi_{0^+}(t+i\epsilon)}
{(t-q^2)^2}
\ ,\label{eq:disrel}
\eeq
where the imaginary part of  $\Pi_{0^+}$ is essentially obtained by
inserting a complete set of hadronic states between the vector
currents in \eq{eq:2pt}:
$$
\cdots q^\mu q^\nu) \mbox{Im}\Pi_{0^+}(q^2+i\epsilon)
=
$$
\beq
\frac{1}{2}\sum_\Gamma (2\pi)^4\delta^{(4)}\l(q-p_\Gamma\r)
\la 0|V^\mu|\Gamma\ra\la\Gamma|V^{\nu\dagger}|0\ra
\ .\label{eq:specfn}
\eeq
The scalar states which couple to $V^\mu$ are $|\Gamma\ra =$
$|B_c(0^+)\ra$, $|B_c\pi\ra$, $\cdots$, $|BD\ra$, $\cdots$, of which
$|BD\ra$ is the state of interest. The obvious positivity of the
spectral function of \eq{eq:specfn} enables one to write an upper
bound on the weighted integral of the form factor $f_0$ along the cut
induced by $|BD\ra$ in $\Pi_{0^+}$:
\beq \chi_{0^+}(q^2) \ge \int_{-\infty}^{-1}\, dw\,
k(w,q^2)\l|f_0(w)\r|^2 \ ,\label{eq:dispbnd}
\eeq
where $k(w,q^2)$ is a kinematical function and where
$w=(M_B^2+M_D^2-t)/(2M_BM_D)$. This bound is interesting because the
function $\chi_{0^+}(q^2)$ can be computed in QCD in terms of an
expansion in $\alpha_s$ and condensates, for $q^2\ll m_b^2$ (i.e. well
below the lowest $|B_c(0^+)\ra$ state). Of course, the more states
$|\Gamma\ra$ one can include in \eq{eq:specfn} in a model-independent
way, the better the bounds on the form factor $f_0$. In particular,
one can use heavy-quark-spin symmetry to include the contributions of
$|B^*D\ra$, $|BD^*\ra$\footnote{$|BD^*\ra$ does not actually
  contribute to the $0^+$ channel.} and $|B^*D^*\ra$, in the semileptonic
domain.

Translating the bound of \eq{eq:dispbnd} into a bound on $f_0$
in the semileptonic region is an exercise in complex analysis. For
convenience, one performs the conformal transformation
\beq
z(w,a) = \frac{\sqrt{w+1}-\sqrt{2}a}{\sqrt{w+1}+\sqrt{2}a}\,;\quad
a>0
\ ,\label{eq:confmap}
\eeq
which maps the cut plane in $w$ onto the unit disc\footnote{For $a=1$, the
zero-recoil point $w=1$ gets mapped onto the origin $z=0$.}. 
Then the bound of \eq{eq:dispbnd} becomes
\beq \chi_{0^+}(q^2) \ge \oint \frac{dz}{2\pi i z}\,|\phi(z,q^2)
f_0(z)|^2 \ ,
\label{eq:confmapbnd}
\eeq
where $\phi(z,q^2)$ is the conformally mapped version of $k(w,q^2)$,
with kinematical singularities inside the unit disc removed.
This bound implies:

$\bullet$ elliptic constraints on the slope, curvature and higher
derivatives of $f_0(z)$ at $z=0$; we show such constraints on
$f_+(z)$, related to $f_0(z)$ by heavy-quark-spin symmetry near $z=0$,
in \fig{fig:ellcst};

$\bullet$ that the remainder of the expansion of $\phi f_0(z)$ in $z$,
once subthreshold singularities are accounted for, can be bounded and
that the expansion converges rapidly, as was first remarked in
  \cite{Boyd:1995cf}.

\FIGURE{\epsfxsize=6cm\epsffile{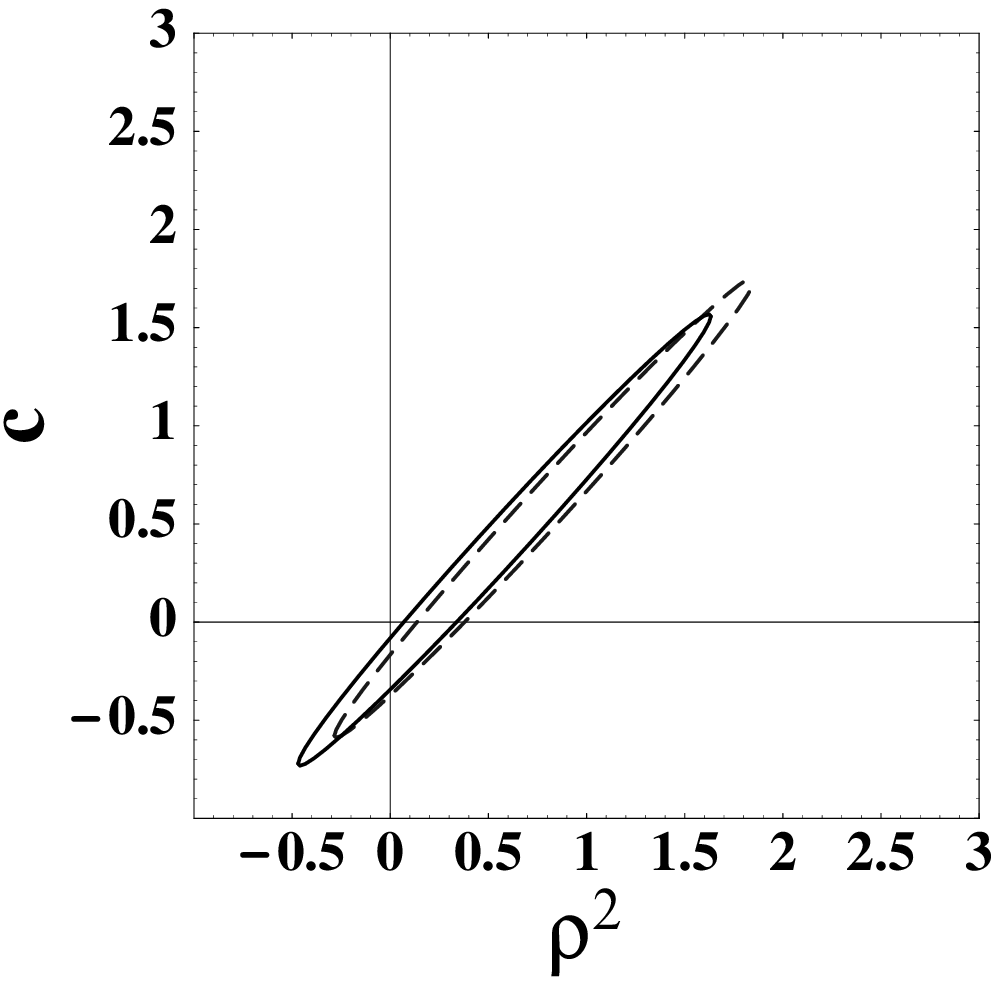}
\caption{Bounds on the parameters $\rho^2$ and $c$ which appear in
  $f_+(w)=f_+(1)(1-\rho^2(w-1)+c(w-1)^2+$ $\cdots)$ 
  \cite{Caprini:1997mu}. Only points within the ellipses are allowed.
  The bounds were obtained in the $0^+$ channel. They take into
  account contributions, amongst others, from the $|BD\ra$, $|B^*D\ra$
  and $|B^*D^*\ra$ states, including (solid) and not including
  (dashed) corrections to heavy-quark spin symmetry.}
\label{fig:ellcst}
}

\medskip
The two most complete analyses are those of \cite{Boyd:1997kz} and
\cite{Caprini:1997mu}. In the former, $\Lambda_b\to\Lambda_c\ell\nu$
decays are also analysed.  One of the upshot of these analyses is that
the form factors for $B^{(*)}\to D^{(*)}$ vector and axial transitions
can each be described with an accuracy better than 2\% with one
parameter and an overall normalisation.

For instance, \cite{Caprini:1997mu} gives for the form factor which determines 
$B\to D\ell\nu$ decays,
$$ \frac{\mathcal{F}_D(w)}{\mathcal{F}_D(1)} =\frac{f_+(w)}{f_+(1)}
\approx 1 - 8 {\rho_1^2}
z + (51.{\rho^2} - 10.) z^2
$$
\beq
 - (252.{\rho^2}-84.) z^3
\ ,
\eeq
with $-0.17<\rho^2<1.51$ and
$z=(\sqrt{w+1}-\sqrt{2})/(\sqrt{w+1}+\sqrt{2})$. The bounds on
$\rho^2$ are slighlty stronger than those given in \fig{fig:ellcst}
as detailed in \cite{Caprini:1997mu}. Similar
parameterisations were given for $\mathcal{F}_{D^*}(w)$.

These dispersive parameterisations have been tried by experimental
collaborations \cite{Bartelt:1999ky} and describe the data very well.
They eliminate the uncertainty in the extrapolation of the
differential decay from $w\ne 1$ to the zero-recoil point.

\subsection{$B\to\pi\ell\nu$ and related work}
\label{sec:bpidisp}

The first application of dispersive bounds to $B\to\pi\ell\nu$ was
performed in \cite{Boyd:1995tt} where it was argued that these
techniques could be used to eliminate certain models for the relevant
form factors. In \cite{Lellouch:1996yv}, these techniques were
extended to extrapolate, to the full kinematical range, lattice
results for $B\to\pi\ell\nu$ form factors obtained at high $q^2$. The
resulting lattice-constrained bounds are shown in \fig{fig:bpibnds},
along with LCSR results. The agreement is excellent.
\FIGURE{
\epsfxsize=6.5cm\epsffile[100 210 530 620]{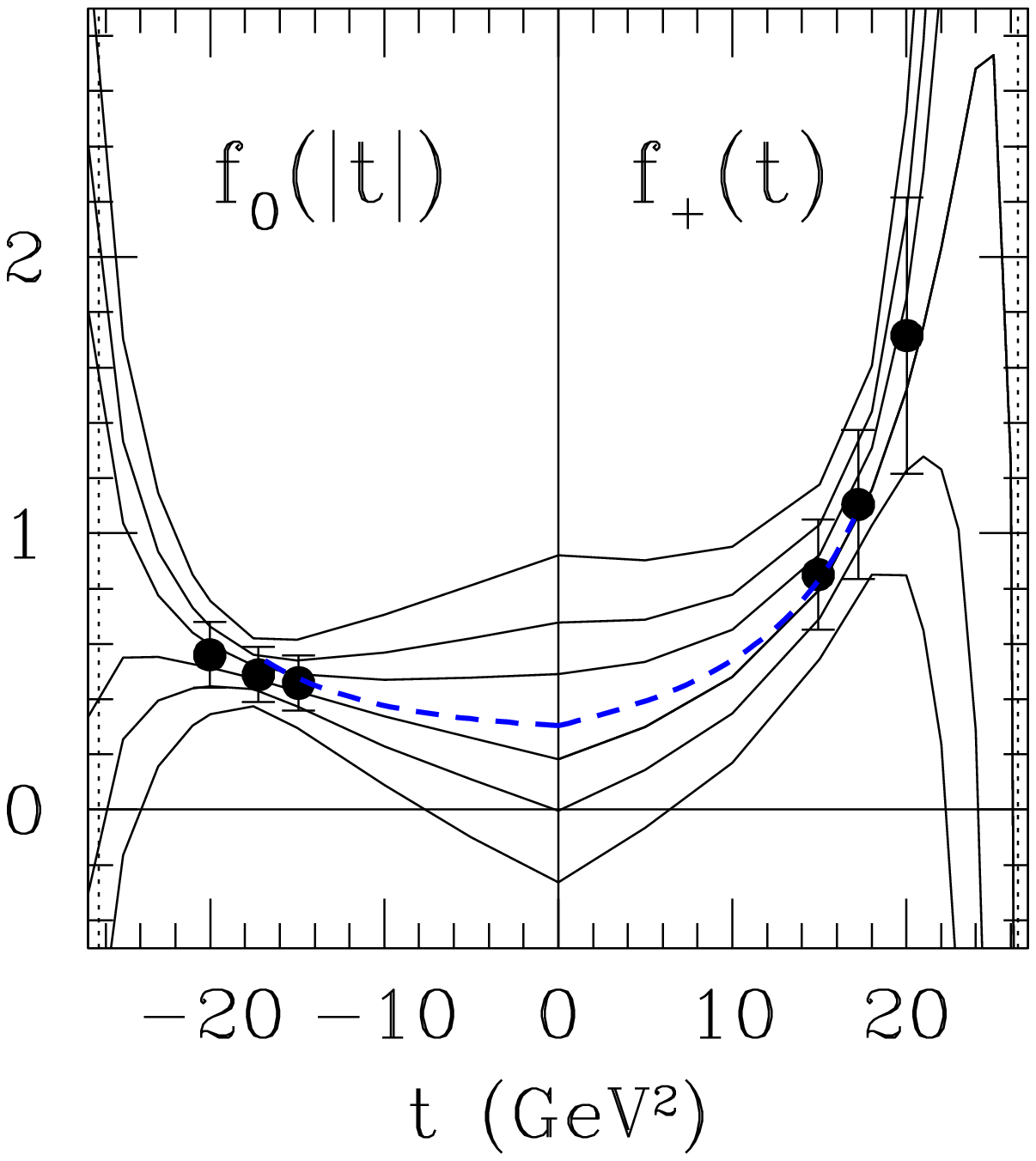}
\caption{Dispersive bounds for $f_0(|t|)$ and $f_+(t)$ 
  in $B^0\to\pi^-\ell^+\nu$ decays \cite{Lellouch:1996yv}.  The
  points are the lattice results of \cite{Burford:1995fc} with added
  systematic errors. The pairs of fine curves are, from the outermost
  to the innermost, the 95\%, 70\% and 30\% bounds, where percentages
  represent the likelihood that the form factor take a value between
  the corresponding pair of curves at the given $t$. The dashed curves
  are the LCSR results of \cite{Ball:1998tj}.}
\label{fig:bpibnds}
}

Similar bounds were obtained for the total rate and are summarized in
\tab{tab:rate}. Since the systematic uncertainties added to the
lattice results of \cite{Burford:1995fc} were generous, the 50\%
bounds may be a reasonable representation of the
constraints given by the dispersive method.
\TABLE{
\begin{tabular}{ccc}
\hline
$\Gamma\l(B^0\to\pi^-\ell^+\nu\r)$ & $f_+(0)$ & CL\\
\hline
$4.8\to 10$ & $0.18\to 0.49$ & 30\% \\
$4.4\to 13$ & $0.10\to 0.57$ & 50\% \\
$3.6\to 17$ & $0.00\to 0.68$ & 70\% \\
$2.4\to 28$ & $-0.26\to 0.92$ & 95\% \\
\hline
\end{tabular}
\caption{Bounds on the rate in units of 
  $|V_{ub}|^2\,ps^{-1}$ and on $f_+(0)$ \cite{Lellouch:1996yv}.}
\label{tab:rate}
}

The results of \cite{Lellouch:1996yv} are rather old and can
presumably be improved with the lattice results and additional
constraints.  For instance, the authors of \cite{Mannel:1998kp}
discuss the inclusion of soft-pion constraints. The authors of
\cite{Boyd:1997qw} discuss the improved constraints brought about by
considering higher moments of the dispersion relations discussed
above, as well as the inclusion of the $B^*\to\pi\ell\nu$ through the
use of heavy-quark spin symmetry.  

The application of these techniques to $B\to\rho\ell\nu$ and $B\to
K^*\gamma$ decays is investigated in \cite{Becirevic:1996ki} and
\cite{Becirevic:1997pd}, respectively.

\section{Conclusion}

\subsection{Lattice}

Many new lattice results for semileptonic decays of the $B$ have or
are about to appear. They have significantly higher statistics than
first generation calculations. This additional statistical accuracy
has enabled some groups to perform the difficult extrapolation to the
physical pion mass in $B\to\pi\ell\nu$ decays, and reconstruct the
$q^2$-behaviour of the physical form factors without $SU(3)$-flavour
assumptions, for $q^2\gsim$ $15\gev^2$.  Moreover, some of these
calculations are non-perturbatively $\ord{a}$-improved indicating that
they have reduced discretisation and matching errors.  Others have
been performed for many lattice spacings, which will enable
extrapolations to the physical continuum or, at least, a better
quantification of discretisation errors. Most of these results,
unfortunately, are still obtained in the unphysical quenched
approximation. However, a number of groups are beginning to partially
include the effect of fermion loops on these decays and preliminary
results were presented at Lattice 99. Furthermore, more and more
theoretical constraints on the $q^2$-behaviour of the relevant form
factors, such as those described in \sec{sec:bpi}, are being taken
into account. This should enable, an extension of the kinematical
range covered by lattice calculations.

Many new results are also appearing for semi\-leptonic $D$ decays
\cite{Simone:1998ti,Maynard:1998ww,Bowler:1999tx,Ryan:1999kx,Abada:1999xd}.
These enable a calibration of the lattice method through comparison
with experiment.  As suggested in \cite{Simone:1998ti}, the ratio of
partial widths for $D\to\pi\ell\nu$ and $D\to K\ell\nu$ is also a good
way to reduce the uncertainty on $|V_{cd}|/$ $|V_{cs}|$, currently
$\sim 17\%$.  Furthermore, the ratio of differential decay rates for
$B\to\rho(\pi)\ell\nu$ to those for the corresponding $D$ decays may
provide a means of determining $|V_{ub}|$ with reduced theoretical
uncertainties \cite{Ryan:1999kx}.

\subsection{Dispersive bounds}

Dispersive bounds for heavy-to-light decays have not been worked on in
a while and can presumably be improved with new lattice results and
additional constraints.

For $B\to D^{(*)}\ell\nu$ decays, dispersive bounds are quite mature
and improvement on current results seems difficult.

\acknowledgments

Warm thanks to L.\ Del Debbio, J.\ Flynn, V.\ Lesk, C.\ Maynard, J.\ 
Nieves and other colleagues of UKQCD for fruitful collaborations and
discussions. I also thank D.\ Becirevic, C.\ DeTar, G.\ Douglas, S.\ 
Hashimoto, J.\ Hein, A.\ Kronfeld, T.\ Onogi and S.\ Ryan for their
help in putting together this review.


\end{document}